\documentclass[a4paper,11pt]{article}
\pdfoutput=1 

\usepackage{jheppub} 

\usepackage{fancyhdr}
\usepackage{graphicx}

\usepackage{braket}
\usepackage{slashed}

\newcommand{\be}{\begin{equation}}
\newcommand{\ee}{\end{equation}}
\newcommand{\bea}{\begin{eqnarray}}

\newcommand{\eea}{\end{eqnarray}}

\newcommand{\Rmnum}[1]{\expandafter\@slowromancap\romannumeral #1@}

\title{\boldmath {\it Fermi}-LAT $\gamma$-ray anisotropy and intensity explained by unresolved Radio-Loud Active Galactic Nuclei}

\author[a,b,c]{Mattia Di Mauro,}
\author[a,b]{Alessandro Cuoco,}
\author[a,b]{Fiorenza Donato}
\author[d]{Jennifer M.~Siegal-Gaskins}

\affiliation[a]{Dipartimento di Fisica, Universit\`a di Torino, via P. Giuria 1, 10125 Torino, Italy}
\affiliation[b]{Istituto Nazionale di Fisica Nucleare, Sezione di Torino, Via P. Giuria 1, 10125 Torino, Italy}
\affiliation[c]{LAPTh, Universit\'e de Savoie, CNRS, 9 Chemin de Bellevue, B.P.\ 110, F-74941 Annecy-le-Vieux, France}
\affiliation[d]{California Institute of Technology, 1200 E. California Blvd., Pasadena, CA 91125, USA}
\emailAdd{mattia.dimauro@to.infn.it}
\emailAdd{alessandro.cuoco@to.infn.it}
\emailAdd{donato@to.infn.it}
\emailAdd{jsg@tapir.caltech.edu}

\abstract{Radio-loud active galactic nuclei (AGN) are expected to contribute substantially to both the intensity and anisotropy of the isotropic $\gamma$-ray background (IGRB).  In turn, the measured 
properties of the IGRB can be used to constrain the characteristics of proposed contributing source classes.  We consider individual subclasses of radio-loud AGN, including low-, intermediate-, and 
high-synchrotron-peaked BL Lacertae objects, flat-spectrum radio quasars, and misaligned AGN.  Using updated models of the $\gamma$-ray luminosity functions of these populations, we evaluate 
the energy-dependent contribution of each source class to the intensity and anisotropy of the IGRB\@.  We find that collectively radio-loud AGN can account for the entirety of the IGRB intensity and 
anisotropy as measured by the {\it Fermi}  Large Area Telescope (LAT).  Misaligned AGN provide the bulk of the measured intensity but a negligible contribution to the anisotropy, while high-synchrotron-peaked BL Lacertae objects provide the dominant contribution to the anisotropy.  In anticipation of upcoming measurements with the {\it Fermi}-LAT and the forthcoming Cherenkov 
Telescope Array, we predict the anisotropy in the broader energy range that will be accessible to future observations. }
\begin{document} 
\maketitle
\flushbottom

\section{Introduction}\label{sec:int}
The intensity and anisotropy of a  diffuse $\gamma$-ray  
background encode information about its contributing sources. The isotropic $\gamma$-ray background (IGRB) is 
the diffuse residual $\gamma$-ray emission, apparent especially at high Galactic latitudes,
observed when the Galactic diffuse emission is subtracted from the observed $\gamma$-ray sky, and when resolved
point sources are either subtracted or masked.
The origin of this emission is not yet fully understood,
but it is thought to 
originate from unresolved sources of extragalactic and possibly Galactic origin.  Recent measurements of the intensity~\cite{Abdo:2010nz} and angular power 
spectrum~\cite{2012PhRvD..85h3007A} of the IGRB by the {\it Fermi} Large Area Telescope (LAT)  have enabled more detailed studies of the contributors to this emission.

The intensity spectrum of the IGRB is largely consistent with a single power law in the energy range of 200~MeV -- 100~GeV~\cite{2012PhRvD..85h3007A}. 
 However, it is expected that many $\gamma$-ray source classes contribute to the IGRB over this energy range, including $\gamma$-ray emitting classes of active galactic nuclei (AGN)~\cite{DiMauro:2013xta,Ajello:2011zi,DiMauro:2013zfa,Ajello:2013lka,2011ApJ...733...66I,1996ApJ...464..600S, 2007ApJ...659..958D,2009ApJ...702..523I}, star-forming galaxies~\cite{Ackermann:2012vca,2010ApJ...722L.199F,Tamborra:2014xia,2014ApJ...786...40L,Thompson:2006qd}, Galactic millisecond 
 pulsars (MSPs)~\cite{Calore:2014oga,SiegalGaskins:2010mp,2013AA...554A..62G}, as well as proposed source classes such as annihilating or decaying dark matter~\cite{Calore:2013yia,Cirelli:2009dv,Abazajian:2010zb,Zavala:2009zr,Cholis:2013ena}. 
 
The {\it Fermi}-LAT, using approximately 2 years of data,  has measured the angular power spectrum of the 
diffuse emission at Galactic latitudes $|b| > 30^{\circ}$, in four energy bins spanning 1 to 50 GeV~\cite{2012PhRvD..85h3007A}.
At multipoles $l \geq 155$, an angular power  above the photon noise level is detected at $> 99.99\%$ CL in the 1-2 GeV, 2-5 GeV, and 5-10 GeV energy bins, 
and at $> 99\%$ CL in the 10-50 GeV energy range. 
Within each energy bin, the measured angular power takes approximately 
the same value at all multipoles, suggesting that it originates from the contribution of one or more unclustered point source populations. We denote this multipole-independent 
anisotropy as a function of energy $C_{\rm P}(E)$. 

In this work we predict both the intensity of the anisotropy and its energy dependence according to the most recent $\gamma$-ray 
emission models of radio-loud AGN, and compare the results to the {\it Fermi}-LAT data. 
Radio-loud AGN sources are a small fraction of AGN (15-20 $\%$) but are the most powerful ones, 
with a ratio of radio (at 5 GHz) to optical (B-band) flux greater than 10~\cite{1989AJ.....98.1195K}.
This category of AGN is divided into blazars or misaligned (MAGN) sources according to the angle of the jets with respect to the line of sight (los). 
Blazars (MAGN) are objects with an emission angle smaller (larger) than about $14^{\circ}$~\cite{urry1995}.
Furthermore, blazars are traditionally divided into flat spectrum radio quasars (FSRQs) and 
BL Lacertae (BL Lac) objects according to the presence or absence of strong broad emission lines 
in their optical/UV spectrum, respectively~\cite{1980ARAA..18..321A,urry1995}.
Extending the classification proposed for BL Lacs~\cite{1995ApJ...444..567P}, all blazars could also be divided according to the value of the synchrotron-peak frequency 
$\nu_{\rm S}$ of their spectrum. The low-synchrotron-peaked (LSP) blazars have the observed peak frequency in the far-infrared (IR) or IR band ($\nu_{\rm S} < 10^{14}$ Hz), 
intermediate-synchrotron-peaked (ISP) blazars have $\nu_{\rm S}$ in the near-IR to ultraviolet (UV) frequencies ($10^{14}$ Hz $< \nu_{\rm S} < 10^{15}$ Hz), while for 
high-synchrotron-peaked (HSP) blazars  the peak frequency is located in the UV band or at higher energies ($\nu_{\rm S} > 10^{15}$ Hz)~\cite{2010ApJ...716...30A}. 
This classification is relevant also for $\gamma$-ray energies because the shape and the intensity of the spectral energy distribution at such high energies is connected to the 
position of the synchrotron peak: the smaller (larger) the $\nu_{\rm S}$, the softer (harder) the $\gamma$-ray photon index $\Gamma$, and the larger (smaller) the $\gamma$-ray flux~\cite{2011ApJ...743..171A,1998MNRAS.299..433F}.

Radio-loud AGN are the most numerous population in the {\it Fermi}-LAT catalogs~\cite{Abdo:2010ru,Fermi-LAT:2011iqa,TheFermi-LAT:2013xza}. 
Previous works have derived the redshift $z$, $\gamma$-ray luminosity $L_{\gamma}$, and photon index $\Gamma$ 
distributions for the detected sources together with predictions for the $\gamma$-ray flux from the unresolved component~\cite{DiMauro:2013xta,Ajello:2011zi,DiMauro:2013zfa,Ajello:2013lka}.
Some of the main results of those studies, which will be used in the present work, are summarized below:
\begin{enumerate}
\item In~\cite{DiMauro:2013xta} the $\gamma$-ray emission from the MAGN population was predicted using a sample of sources detected in $\gamma$-rays and calibrated using radio data
in order to construct the $\gamma$-ray luminosity function. 
These sources have a mean  photon index of $2.37\pm0.32$ and a $\gamma$-ray luminosity which is 
about two orders of magnitude larger than the radio core luminosity at 5 GHz. 
The best-fit value of the unresolved emission from MAGN was found to be 25-30\% of the
IGRB for $E > 100$~MeV, enveloped in an uncertainty band of about a factor of ten. 
\item The FSRQ population was analyzed in~\cite{Ajello:2011zi}, where it was found that 
FSRQ objects are nearly all LSP blazars, with a broad redshift distribution
 spanning from 0.2 to 3 and a mean  photon index of $2.44\pm0.18$.
FSRQs are powerful sources with the high-energy peak of the spectral energy distribution (SED) in the range of 10~MeV -- 1~GeV.
The unresolved emission from this component contributes $9.3^{+1.6}_{-1.0}\%$ to the IGRB for $E > 100$~MeV, 
with a steeply falling spectrum at energies above $\sim$~5-10~GeV.
\item In~\cite{DiMauro:2013zfa,Ajello:2013lka} the population of BL Lacs was studied 
in terms of the redshift, $\gamma$-ray luminosity, and photon index distributions. 
In particular,~\cite{DiMauro:2013zfa} studied the SED and $\gamma$-ray luminosity function separately for the LSP/ISP/HSP BL Lacs
using also the high-energy $\gamma$-ray spectra measured by Imaging Atmospheric Cherenkov Telescopes (IACTs).
LSP and ISP BL Lacs are found to be statistically the same $\gamma$-ray population with a mean  photon index of $2.08\pm0.15$ 
and an exponential cut-off at $37^{+85}_{-20}$ GeV, hence they are associated to a unique class called LISP (LSP+ISP). HSPs have a mean  
photon index of $1.86\pm0.16$ with an exponential cut-off at $910^{+1100}_{-450}$ GeV.
The $\gamma$-ray emission from unresolved BL Lac sources was derived in~\cite{DiMauro:2013zfa,Ajello:2013lka} to be about $7-11\%$ of the IGRB in the 
range 100~MeV -- 1~GeV, and as much as 100\% for energies higher than 100~GeV.
\end{enumerate}

While the anisotropy spectrum is a relatively recently available observable, historically the 
diffuse extragalactic $\gamma$-ray sky has been studied through the energy spectrum of the IGRB\@.
Measurements of the spectrum of the IGRB  in the energy range 200~MeV -- 100~GeV have been reported by the  {\it Fermi}-LAT Collaboration in~\cite{Abdo:2010nz} for $b>10^{\circ}$. 
More recently the {\it Fermi}-LAT Collaboration has presented preliminary results for the IGRB spectrum
 in the broader energy range of 100~MeV -- 820~GeV~\cite{bechtol2014}. 
In~\cite{DiMauro:2013zfa,Cholis:2013ena} it was shown that it is possible to explain the entire spectrum of the IGRB 
by the unresolved emission from the FSRQ, BL Lac, MAGN, MSP, and star-forming galaxy populations.

The information available from the anisotropy measured  in~\cite{2012PhRvD..85h3007A} 
has been used, for example, in~\cite{Cuoco:2012yf} together with the
source count distribution of blazars to show that they  contribute  only by about 20-30\% to the IGRB \emph{intensity}, confirming with the anisotropy the result found via the source counts alone~\cite{Collaboration:2010gqa}.  Anisotropy from blazars has been further studied in~\cite{Harding:2012gk,Chang:2013hia}.  It has also 
been used to constrain the contribution of MSPs to the IGRB~\cite{SiegalGaskins:2010mp}
showing that stronger constraints are obtained with respect to  the case when intensity alone is used. A recent analysis 
has demonstrated that these galactic sources contribute indeed negligibly to the measured anisotropy, as well as to the 
IGRB intensity~\cite{Calore:2014oga}.
The anisotropy of star-forming galaxies has been studied in~\cite{Ando:2009nk}.
Finally, several works have investigated the anisotropy  from dark matter annihilation into $\gamma$-rays~\cite{2006PhRvD..73b3521A,2007PhRvD..75f3519A,SiegalGaskins:2008ge,Hensley:2009gh,2009PhRvD..80b3520A,2009PhRvD..80b3518F,2009PhRvL.102x1301S,2011MNRAS.414.2040C,Calore:2014hna}.

In this work we compare radio-loud AGN model predictions to both
the intensity and the anisotropy of the IGRB
and we will show that  a coherent picture can be constructed in which radio-loud AGN account for the measured values of both of these observables.
This is the first time that an attempt to simultaneously explain the $\gamma$-ray flux and anisotropy data 
has been pursued using a single underlying global model of the unresolved emission.
In \S\ref{sec:model} we describe the models for the radio-loud AGN populations and present the calculation of their intensity and anisotropy contributions to the IGRB\@.  We discuss the results and 
compare them to the measured intensity and anisotropy by the {\it Fermi}-LAT in \S\ref{sec:results}.
In addition to comparing model predictions for the intensity and anisotropy of the IGRB in the energy range 100~MeV -- 100~GeV, relevant for {\it Fermi}-LAT observations,
we also study the energy range 100~GeV--10~TeV, as will be covered 
by the forthcoming Cherenkov Telescope Array (CTA) observatory~\cite{Consortium:2010bc,Dubus:2012hm}.
In \S\ref{sec:cta} we  derive the expected angular power from unresolved radio-loud AGN 
in the higher energy range relevant for future CTA observations and compare it with the expected sensitivity reach of CTA\@.  We discuss and conclude in \S\ref{sec:conc}.

\section{Anisotropy Model Predictions}
\label{sec:model}

\subsection{The Angular Anisotropy}\label{sec:Cp}
The angular power $C_{\rm P}$ produced by the unresolved flux of an unclustered point source
population is derived using the following equation~\cite{Scott:1998ei,Cuoco:2012yf,2012PhRvD..85h3007A}:
\begin{equation}
     \label{Cpdef}
        C_{\rm P}(E_0 \leq E \leq E_1) = \int_{\Gamma_{\rm min}}^{\Gamma_{\rm max}} d\Gamma \int^{S_{\rm t}(\Gamma)}_0 S^2 \frac{d^2N}{dS d\Gamma} dS,
    \end{equation} 
where $S$ is the photon flux of the source integrated in the range $E_0 \leq E \leq E_1$ in units of ph cm$^{-2}$ s$^{-1}$,  while
$S_{\rm t}(\Gamma)$ denotes the flux detection threshold as function of the photon index of the source $\Gamma$ (see below), 
 and where  $\Gamma_{\rm min}$--$\Gamma_{\rm max}$ is its range of variation. 
Finally, $d^2N/(dSd\Gamma)$ is the differential number of sources per unit flux $S$, unit photon index $\Gamma$ and unit solid angle.

It is well known that a strong bias is present between the flux and the photon index of sources detected by the {\it Fermi}-LAT 
when considering fluxes integrated in the range 100~MeV -- 100~GeV.
Sources with a photon index of 1.5 can be detected to fluxes (100~MeV -- 100~GeV) a factor of about 20 fainter than those at which a source with a photon index of 3.0 can be 
detected~\cite{Fermi-LAT:2011iqa,Abdo:2010ru}.
This means that the function $S_{\rm t}(\Gamma)$ cannot be approximated as a constant in $\Gamma$ when considering fluxes in that energy range.
On the other hand, it has been shown that the bias is almost absent if the fluxes $S$ integrated above 1~GeV are considered~\cite{Cuoco:2012yf,Fermi-LAT:2011iqa}.
In this case, the function $S_{\rm t}(\Gamma)$ can be simply approximated as a constant  $S_{\rm t}(\Gamma)=S_{>1}$,
where $S_{>1}$ is the flux integrated above 1 GeV.
Further, as we will show in \S\ref{sec:fit}, since the source spectra are described by Eqs.~\ref{dNdE1}, \ref{dNdE2}, and \ref{dNdE3}, a relation between $S_{>1}$ and the 
integrated flux between $E_0 $ and  $ E_1$ can easily be found, and thus the function $S_{\rm t}(\Gamma)$ can be calculated for any energy range considered.

Following~\cite{Cuoco:2012yf} we adopt the value $S_{>1}=5\cdot10^{-10}$ ph cm$^{-2}$s$^{-1}$,
which is appropriate when the 1FGL catalogue is used as reference for the resolved point sources.
This is a consistent choice with respect to the {\it Fermi}-LAT anisotropy measurements, which were made after masking the 1FGL sources, and with which we compare the predicted model anisotropy.


\subsection{Luminosity Function and the Source Count Distribution}\label{sec:N}

To derive the $d^2N/(dSd\Gamma)$ required to calculate $C_{\rm P}$  for the MAGN, FSRQ and BL Lac sources,
the quantity we will use for each population is the  $\gamma$-ray luminosity function (LF) 
$\rho_{\gamma}(L_{\gamma},z,\Gamma)=dN/d\Gamma dz dL_{\gamma}$
which specifies the comoving number density of the given objects, differentially 
per rest-frame luminosity $L_{\gamma}$, redshift $z$, and photon index $\Gamma$.  The LF completely characterizes the specified source population.
We will use the LFs of MAGN, FSRQ and BL Lac populations as derived in~\cite{DiMauro:2013xta,Ajello:2011zi,DiMauro:2013zfa}.
In these models, the LF is assumed to be separable in the $\Gamma$ variable,
whose distribution is parameterized as a Gaussian function:
\begin{equation}
     \label{dNdGamma}
	\frac{dN}{d\Gamma} \propto \exp{\left(-\frac{(\Gamma-\bar{\Gamma})^2}{2\sigma^2}\right)},
    \end{equation} 
with the values for the mean spectral index $\bar{\Gamma}$ and the dispersion $\sigma$ fixed, for each population, to the numbers reported in \S\ref{sec:int}.
With a slight abuse of notation we will  also write  $\rho_{\gamma}(L_{\gamma},z,\Gamma)=dN/d\Gamma \, \rho_{\gamma}(L_{\gamma},z)$.

To simplify the discussion, and also to facilitate the comparison with available data,
in this section we consider the cumulative source count distribution $N(>S)$, which represents the number of sources with a flux larger than $S$.
The same methods can be applied to $d^2N/(dSd\Gamma)$, as we briefly
discuss in \S\ref{sec:fit}.
The $N(>S)$ can be obtained from the LF as~\cite{DiMauro:2013xta,Ajello:2013lka,Ajello:2011zi,DiMauro:2013zfa}:
\begin{equation}
     \label{Ncount}
	N(>S) = \Delta \Omega \int^{\Gamma_{\rm max}}_{\Gamma_{\rm min}} d\Gamma \int^{z_{\rm max}}_{z_{\rm min}} dz \int^{L_{\gamma,{\rm max}}}_{L_{\gamma}(S,\Gamma,z)} 
	dL_{\gamma} \frac{dV}{dz} \frac{dN}{d\Gamma} \rho_{\gamma}(L_{\gamma},z),
    \end{equation} 
where $dV/dz$ is the comoving volume per unit redshift~\cite{Hogg:1999ad} and $\Delta \Omega$ is the solid angle. 
We will use in the following a $\Delta \Omega$ of $2\pi$  corresponding to  $\gamma$-ray
sources above a Galactic cut of $\pm$30$^\circ$. This is opposed to another common convention 
where  $N(>S)$ is divided by  $\Delta \Omega$ and expressed in units of deg$^{-2}$.
The limits of integration $\Gamma_{\rm min}$, $\Gamma_{\rm max}$, $z_{\rm min}$, $z_{\rm max}$, and 
$L_{\gamma,{\rm max}}$ are taken from~\cite{DiMauro:2013xta,Ajello:2011zi,DiMauro:2013zfa}, 
although we note that the results depend only weakly on the specific values of the limits.
Finally $L_{\gamma}(S,\Gamma,z)$  represents the rest-frame $\gamma$-ray luminosity
for a source with a photon index $\Gamma$ at redshift $z$ with observed photon flux $S$,
and will be derived in the next section. 
Both $S$ and $L_{\gamma}$ refer to integrated quantities in the relevant energy range (see next section).

\begin{figure}
\centering
\includegraphics[width=0.48\textwidth]{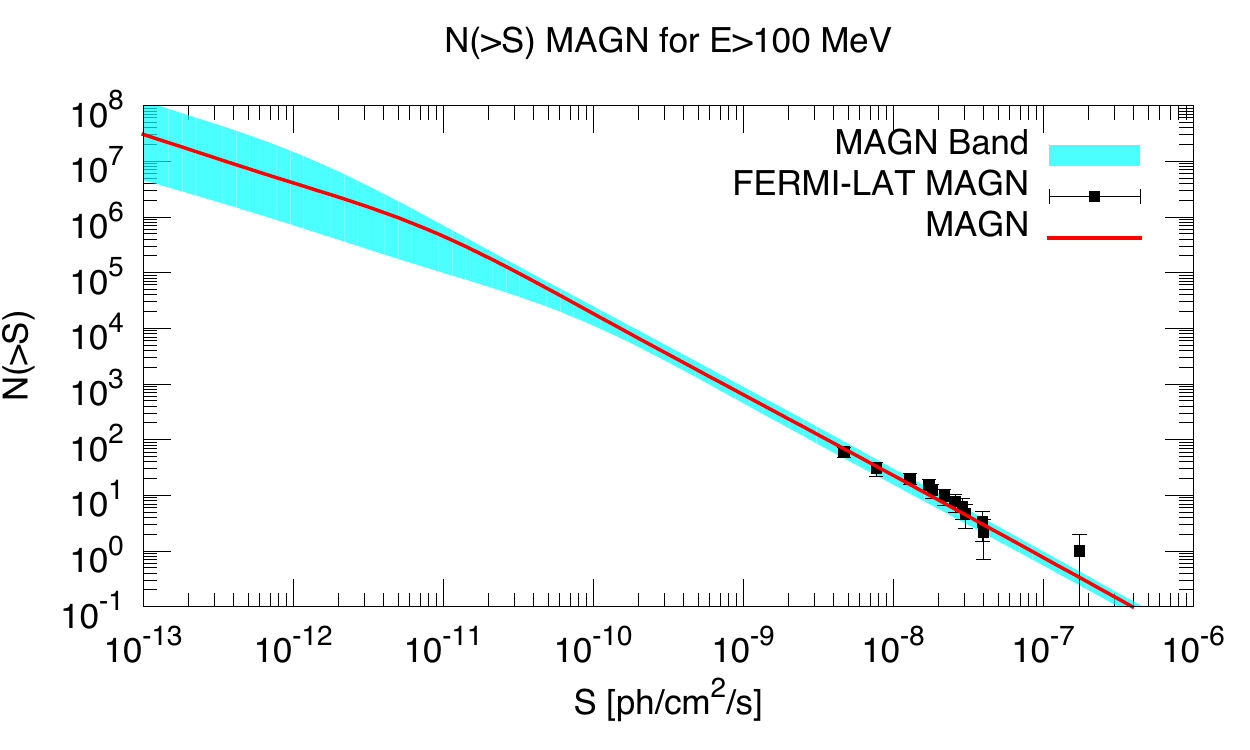}
\includegraphics[width=0.48\textwidth]{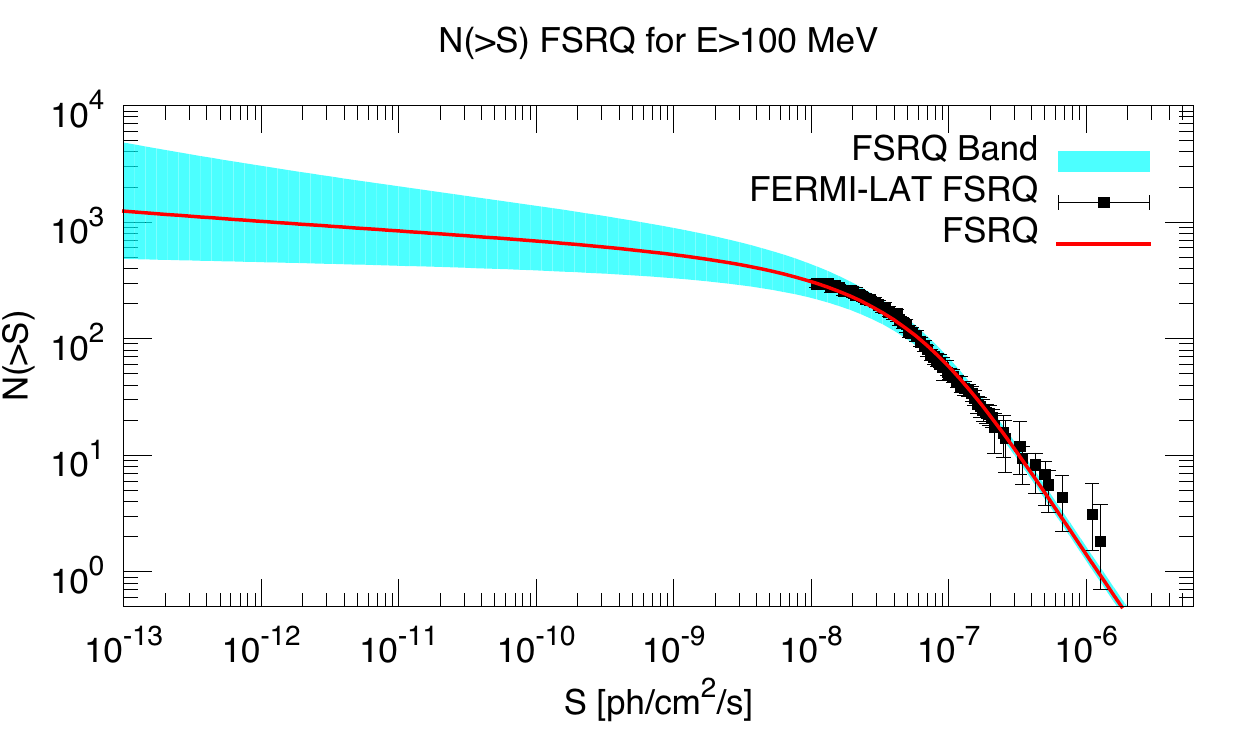}
\includegraphics[width=0.48\textwidth]{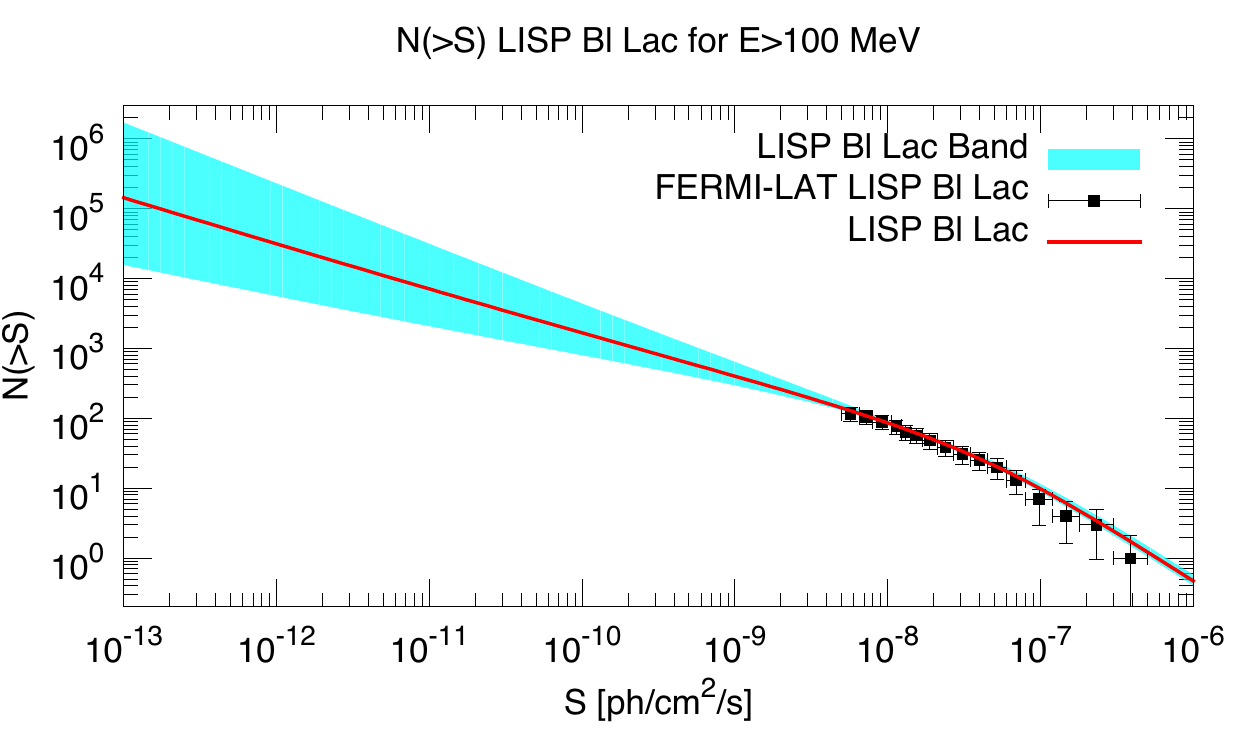}
\includegraphics[width=0.48\textwidth]{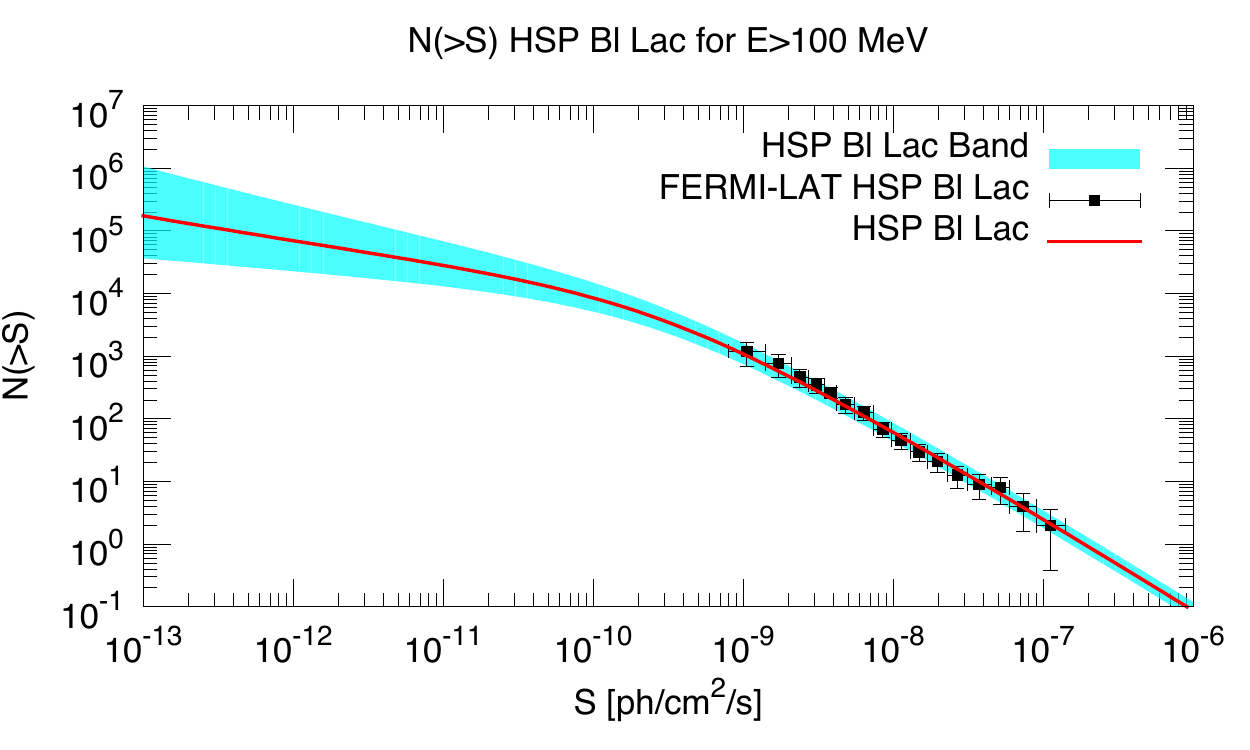}
\caption{The theoretical cumulative source count distribution $N(>S)$ (solid red line) together with the $1\sigma$ uncertainty band (cyan) is shown for radio-loud AGN classes as 
derived in~\cite{DiMauro:2013xta,Ajello:2011zi,DiMauro:2013zfa}: MAGN (top left), FSRQs (top right), LISP (bottom left) and HSP (bottom right) BL Lacs. The data (black points) of the {\it Fermi}-LAT 
experimental source count distribution for the sources of each class are also shown, taken from~\cite{DiMauro:2013xta,Ajello:2011zi,DiMauro:2013zfa}.}
\label{fig:ncount}
\medskip
\end{figure}

In Fig.~\ref{fig:ncount} the theoretical source count distribution $N(>S)$ in terms of $S_{0.1-100 {\rm GeV}}$ is shown, 
together with the $1\sigma$ uncertainty band,  
for MAGN~\cite{DiMauro:2013xta}, FSRQs~\cite{Ajello:2011zi}, and BL Lac LISP (LSP+ISP) and HSP objects~\cite{DiMauro:2013zfa}.
The $1\sigma$ band has been derived for each AGN class considering the uncertainty 
on the $\gamma$-ray emission mechanism and on the redshift, $\gamma$-ray luminosity, and photon index distributions.
In the case of MAGN~\cite{DiMauro:2013xta}, the number of detected sources in the {\it Fermi}-LAT catalogs 
is too small to determine a $\gamma$-ray LF, and a correlation between 
the $\gamma$-ray and radio emission from the core of the MAGN was performed.
The $\gamma$-ray luminosity function was then derived from the radio luminosity function of~\cite{willott}.
The calibration of this correlation is the main source of uncertainty for 
the theoretical prediction of the MAGN source count distribution shown in Fig.~\ref{fig:ncount}.
This uncertainty also leads to a large uncertainty in the prediction of the 
the unresolved $\gamma$-ray emission from MAGN~\cite{DiMauro:2013xta}.
Blazars, FSRQ, and BL Lacs are numerous in the {\it Fermi}-LAT catalogs~\cite{Abdo:2010ru,Fermi-LAT:2011iqa,TheFermi-LAT:2013xza}, hence in~\cite{Ajello:2011zi,DiMauro:2013zfa} the $
\rho_{\gamma}$ was derived directly from $\gamma$-ray data. 
In this case the uncertainties on the source count distributions and on the unresolved $\gamma$-ray emission come from the uncertainty of the redshift, $\gamma$-ray luminosity, and photon index 
distributions, and of the SEDs of these sources. 

In Fig.~\ref{fig:ncount}  we also show the experimental determinations of $N(>S)$
as derived from resolved $\gamma$-ray sources, again, normalized  to the number of sources 
above $\pm$30$^\circ$
in Galactic latitude, coherently with the theoretical model predictions.
The data points have been taken from~\cite{DiMauro:2013xta,Ajello:2011zi,DiMauro:2013zfa}, 
and have been derived as:
\begin{equation}
\label{eq:expNcount}
   \displaystyle 	N_{exp}(>S) = \sum^{N_{S}}_{i=1} \frac{1}{\omega(S_i)},
\end{equation}
where the sum is over the $N_{S}$ sources with a flux $S_i>S$,
and  $\omega(S)$ is the {\it Fermi}-LAT detection efficiency 
for a source with flux $S$ in the energy range 0.1-100~GeV. 
For BL Lacs and MAGN the efficiency $\omega(S)$ derived in~\cite{DiMauro:2013xta} is adopted,
while for FSRQs we refer to the one calculated in~\cite{Collaboration:2010gqa}.
Uncertainties are simply given by the Poissonian errors ($\propto \sqrt(N)$) 
associated with the finite number of sources in each flux bin and by
the uncertainty on the efficiency itself as given in~\citep{Ajello:2011zi,Ajello:2013lka,DiMauro:2013xta}. 
The experimental source count distribution of FSRQs is based on the sources of the 1FGL catalog, 
while that for MAGN is based on both the 1FGL and 2FGL catalogs, and that for BL Lacs on the 2FGL catalog.

\subsection{Energy range rescaling of the cumulative source count distribution}
\label{ncountres}

The source count distributions derived in~\cite{DiMauro:2013xta,Ajello:2011zi,DiMauro:2013zfa} 
and displayed in Fig.~\ref{fig:ncount} are valid in the energy range 100~MeV -- 100~GeV,
while we will need to calculate $C_{\rm P}(E)$ and therefore $d^2N/dSd\Gamma$ in the energy bins [1,10], [1.04,1.99], [1.99,5.00], [5.0,10.4] and [10.4,50.0]~GeV.
Below, we illustrate how we rescale $d^2N/dSd\Gamma$ to the new
energy bands, again demonstrating the procedure on $N(>S)$ rather than $d^2N/dSd\Gamma$ itself.

The source energy spectrum $dN/dE$ is defined by:
\begin{eqnarray}
     \label{dNdE}
	\frac{dN}{dE} (E,\Gamma,E_{\rm c}) &=& K \; \mathcal{F}(E,\Gamma,E_{\rm c}),
    \end{eqnarray} 
where $K$ is a normalization factor and $\mathcal{F}(E,\Gamma,E_{\rm c})$ is the energy-dependent part of the spectrum which depends also on the photon index $\Gamma$ and energy cutoff $E_{\rm c}$. $\mathcal{F}(E,\Gamma,E_{\rm c})$ is given by a simple power law (with $E_{\rm c} \rightarrow \infty$) for MAGN~\cite{DiMauro:2013xta}, 
 a power law with an exponential cut-off  for BL Lacs as in~\cite{DiMauro:2013zfa}, and  a power law with a square-root exponential cut-off 
 for  FSRQs as in~\cite{Ajello:2011zi}:
\begin{eqnarray}
     \label{dNdE1}
	\mathcal{F}_{\rm MAGN}(E,\Gamma) &=& \left( \frac{E}{E_{\rm P}} \right)^{-\Gamma} \\
    \label{dNdE2}
	\mathcal{F}_{\rm BLLAC}(E,\Gamma',E_{\rm c})&=& \left( \frac{E}{E'_{\rm P}} \right)^{-\Gamma'} \exp{\left( -\frac{E}{E_{\rm c}} \right)} \\
   \label{dNdE3}
	\mathcal{F}_{\rm FSRQ}(E,\Gamma'',E'_{\rm c}) &=& \left( \frac{E}{E''_{\rm P}} \right)^{-\Gamma''} \exp{\left(-\sqrt{\frac{E}{E'_{\rm c}}}\right)},
    \end{eqnarray} 
where $E_{\rm c}$ and $E'_{\rm c}$ are the cut-off energies, $E_{\rm P}$, $E'_{\rm P}$ and $E''_{\rm P}$ are the pivot 
energies fixed to 1 GeV and $\Gamma$, $\Gamma'$ and $\Gamma''$ are the photon indexes.
From the full SED given above, the flux $S$ and the $\gamma$-ray luminosity $L_{\gamma}$ in the 
benchmark energy range $E\in [E^{\rm b}_0=0.1,E^{\rm b}_1=100]$~GeV can be calculated as~\cite{Abdo:2010ru,Fermi-LAT:2011iqa}:
\begin{eqnarray}
     \label{S}
	S \equiv S(E^{\rm b}_0 \leq E\leq E^{\rm b}_1) &=& \int^{E^{\rm b}_1}_{E^{\rm b}_0} \frac{dN}{dE} dE, \\
	\label{Lg}
L_{\gamma} \equiv L_{\gamma}(E^{\rm b}_0 \leq E\leq E^{\rm b}_1) &=& 4 \pi d^2_L(z) \int^{E^{\rm b}_1}_{E^{\rm b}_0} \frac{1}{\mathcal{K}(z,\Gamma,E)} \frac{dN}{dE} E dE   =  L_{\gamma}(S,\Gamma,z),
\end{eqnarray} 
where $d_L(z)$ is the luminosity distance and 
$\mathcal{K}(z,\Gamma,E)$ is the K-correction, i.e., the ratio between the observed and the rest-frame luminosity in
the given energy range. For the three SEDs the K-correction can be calculated, respectively, as 
\begin{eqnarray}
     	\label{K1}
	\mathcal{K}(z,\Gamma) &=& (1+z)^{2-\Gamma} \\
	 \label{K2}
	\mathcal{K}(z,\Gamma,E,E_{\rm c}) &=& (1+z)^{2-\Gamma} \exp{\left( -\frac{E z}{E_{\rm c}} \right)}\\
	 \label{K3}
	\mathcal{K}(z,\Gamma,E,E_{\rm c}) &=& (1+z)^{2-\Gamma} \exp{\left(-\frac{(\sqrt{E(1+z)}-\sqrt{E})}{\sqrt{E_{\rm c}}} \right)}.
    \end{eqnarray} 
Using  Eqs. \ref{dNdE1}-\ref{dNdE3}
 and the definitions of $S$ and $L_{\gamma}$ in Eqs.~\ref{S} and \ref{Lg}, we have rescaled the fluxes $S$ and luminosities $L_{\gamma}$ valid for 100~MeV  to 
 100~GeV energy range into the fluxes $S'$ and luminosities $L'_{\gamma}$ integrated in the energy ranges of $C_{\rm P}(E)$. Formally, these relations can be written as:
\begin{eqnarray}
     	\label{S'}
	S'(E_0 \leq E\leq E_1) &=& \left(\frac{S}{\int^{E^{\rm b}_1}_{E^{\rm b}_0} \mathcal{F}(E,\Gamma,E_{\rm c}) \; dE} \right)
	\int^{E_1}_{E_0} \mathcal{F}(E,\Gamma,E_{\rm c}) dE = S'(S,\Gamma,E_{\rm c})\\
	\label{L'}
	L'_{\gamma}(E_0 \leq E\leq E_1) &=& \left(\frac{L_{\gamma}}{\int^{E^{\rm b}_1}_{E^{\rm b}_0}  \frac{1}{\mathcal{K}(z,\Gamma,E)}  E \mathcal{F}(E,\Gamma,E_{\rm c}) dE} \right)
	\int^{E_1}_{E_0}  \frac{ E \; \mathcal{F}(E,\Gamma,E_{\rm c})}{\mathcal{K}(z,\Gamma,E)}   dE.
    \end{eqnarray} 
Given the above definitions (Eqs.~\ref{S'}, \ref{L'}), the source count distribution $N'(>S')$ for the energy range $E\in[E_0,E_1]$ can be expressed as:
\begin{equation}
     \label{N'count}
	N'(>S') = \Delta \Omega \int^{\Gamma_{\rm max}}_{\Gamma_{\rm min}} d\Gamma \int^{z_{\rm max}}_{z_{\rm min}} dz 
	\int^{L_{\gamma,{\rm max}}}_{L_{\gamma}(S(S',\Gamma),\Gamma,z)} 
	dL_{\gamma} \frac{dV}{dz} \frac{dN}{d\Gamma} \rho_{\gamma}(L_{\gamma},z),
    \end{equation}  
where the relation $S(S',\Gamma)$ can be derived from the definition of Eq.~\ref{S'} and depends on the type of energy spectrum used.
The relation $L_{\gamma}(S,\Gamma,z)$ is also spectrum dependent and is given by Eq.~\ref{Lg}.

\begin{figure}
\centering
\includegraphics[width=0.48\textwidth]{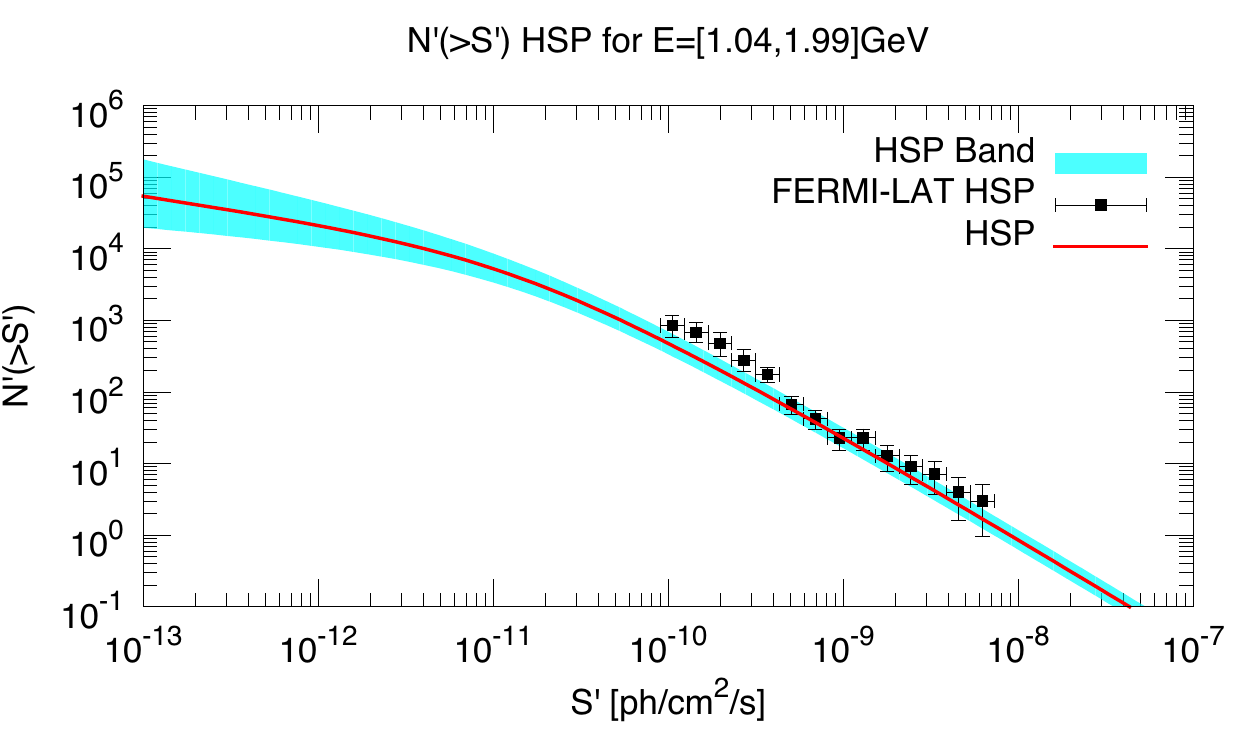}
\includegraphics[width=0.48\textwidth]{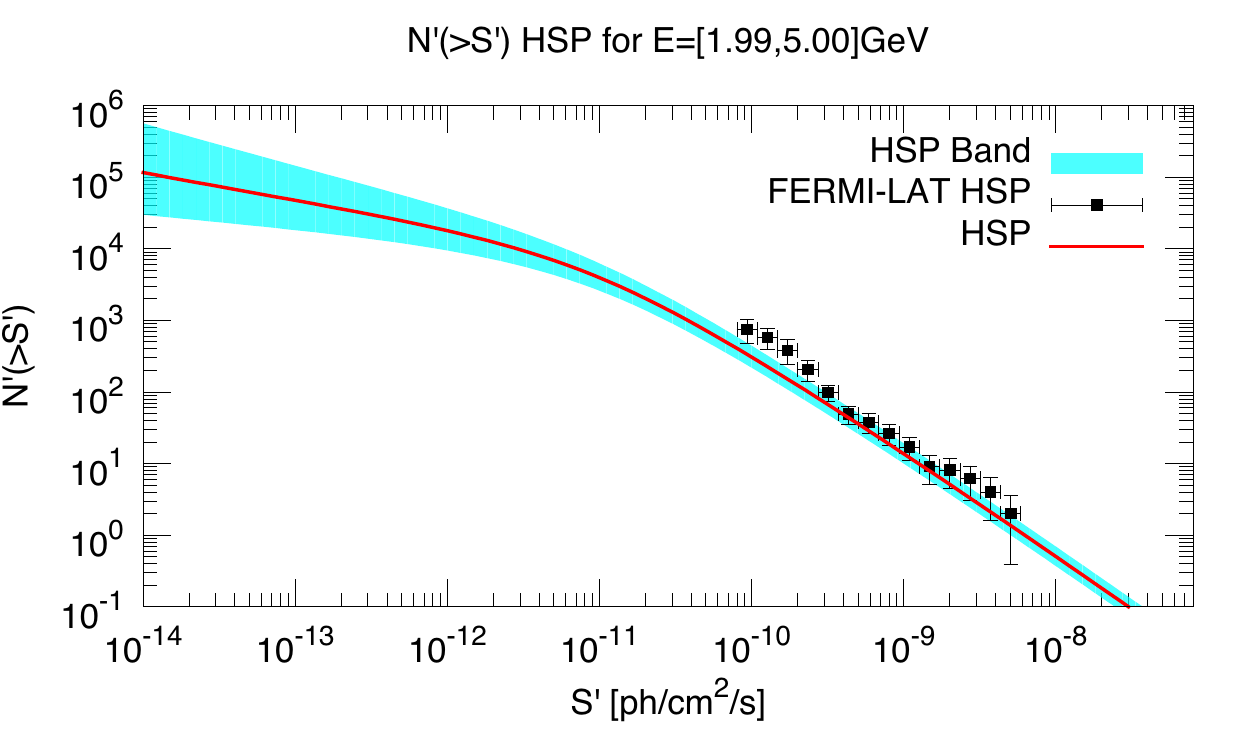}
\includegraphics[width=0.48\textwidth]{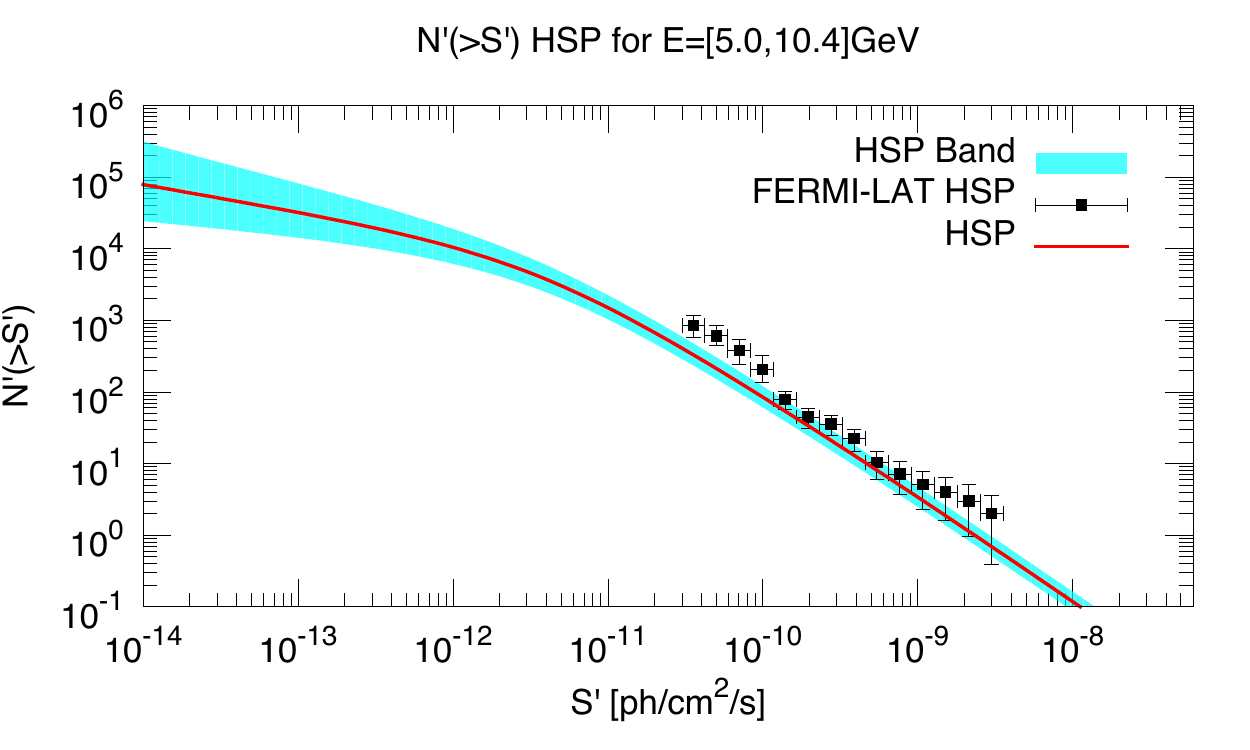}
\includegraphics[width=0.48\textwidth]{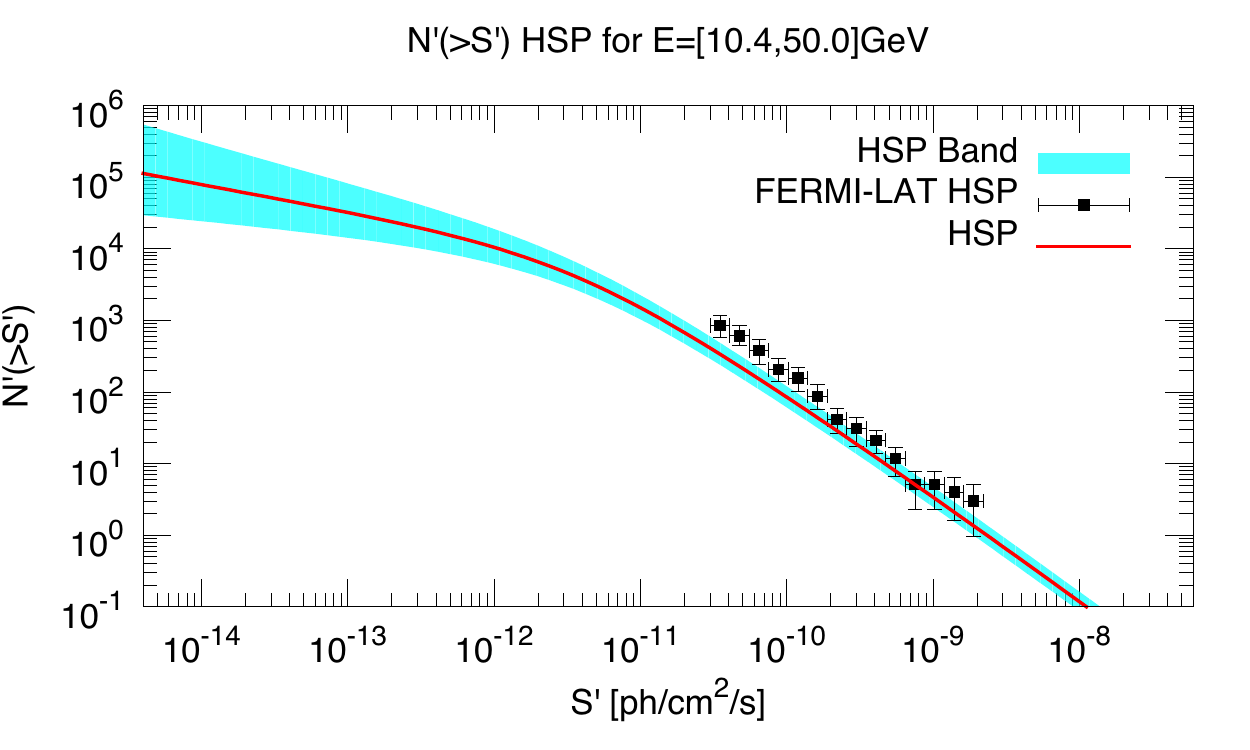}
\caption{The HSP BL Lac theoretical cumulative source count distribution (solid red line) and the 1$\sigma$
 uncertainty band (cyan) are shown for the energy bands [1.04,1.99], [1.99,5.00], [5.0,10.4] and [10.4,50.0]~GeV (clockwise from top left).
We overlay also the {\it Fermi}-LAT experimental data (black points) as derived in~\cite{DiMauro:2013xta,Ajello:2011zi,DiMauro:2013zfa} 
and adapted to the displayed energy bins as described in the text.}
\label{fig:ncounthsp}
\medskip
\end{figure}

\begin{figure}
\centering
\includegraphics[width=0.48\textwidth]{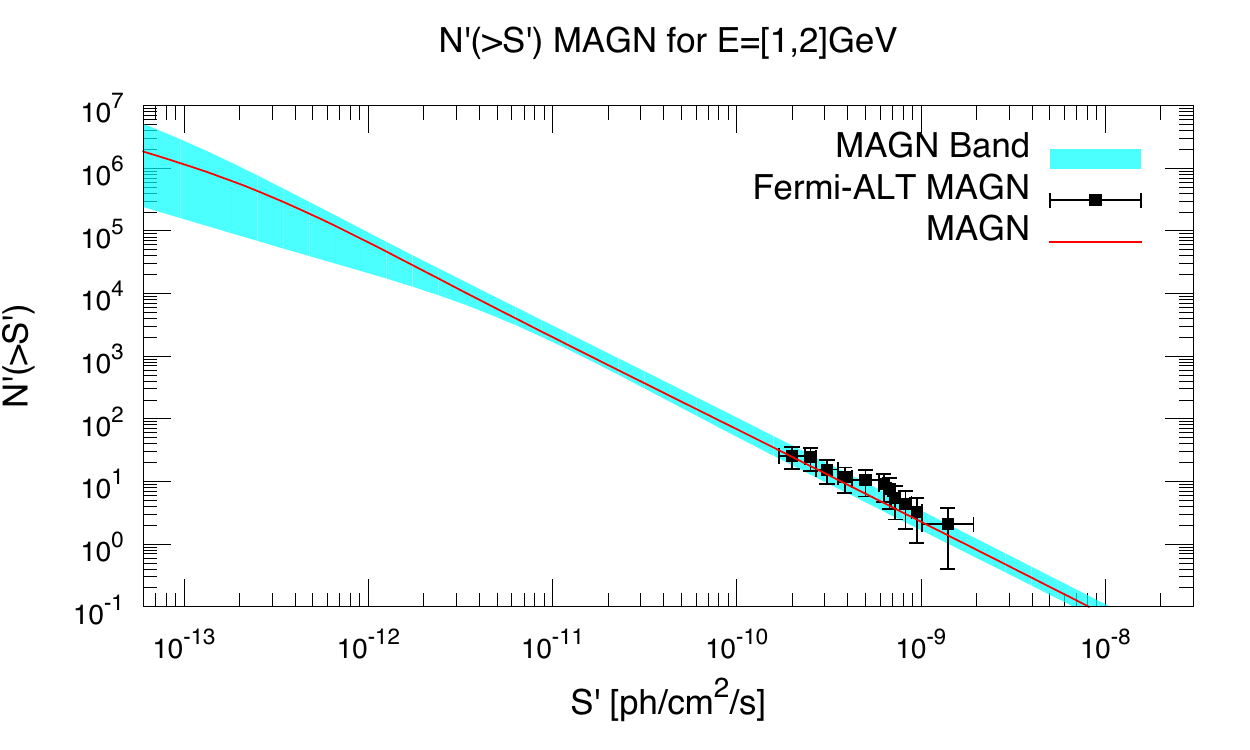}
\includegraphics[width=0.48\textwidth]{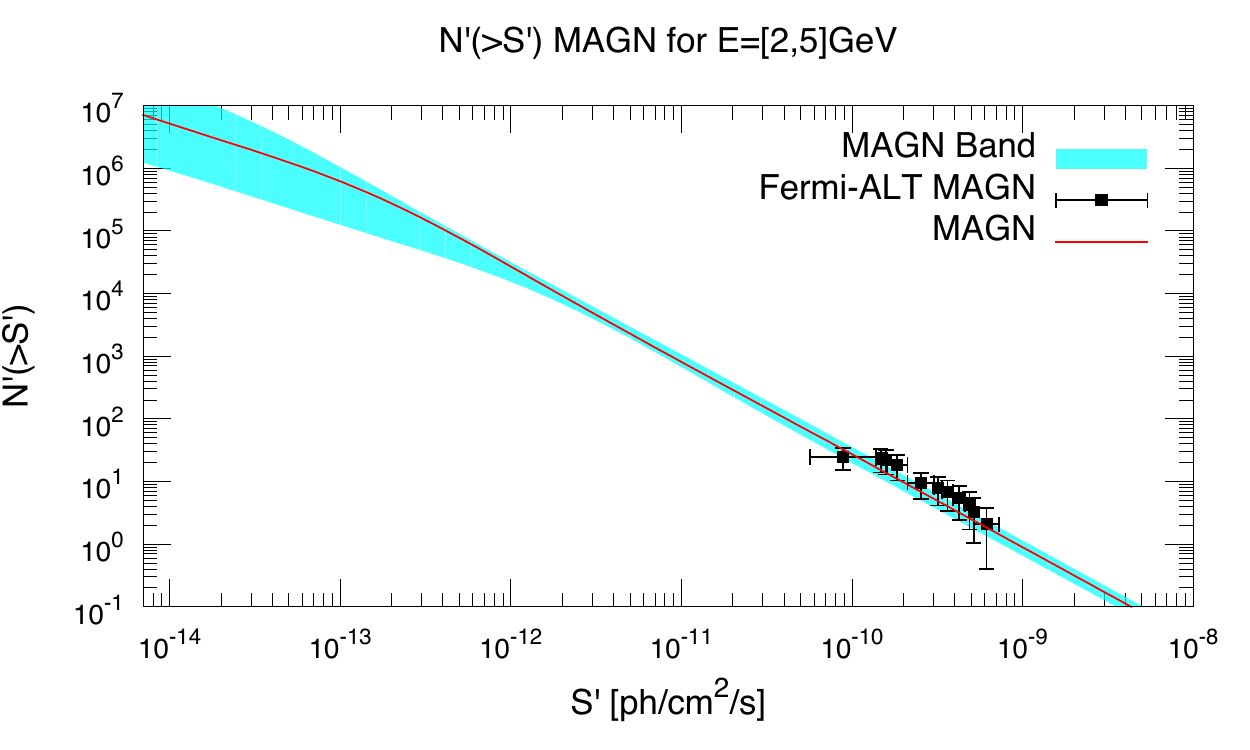}
\includegraphics[width=0.48\textwidth]{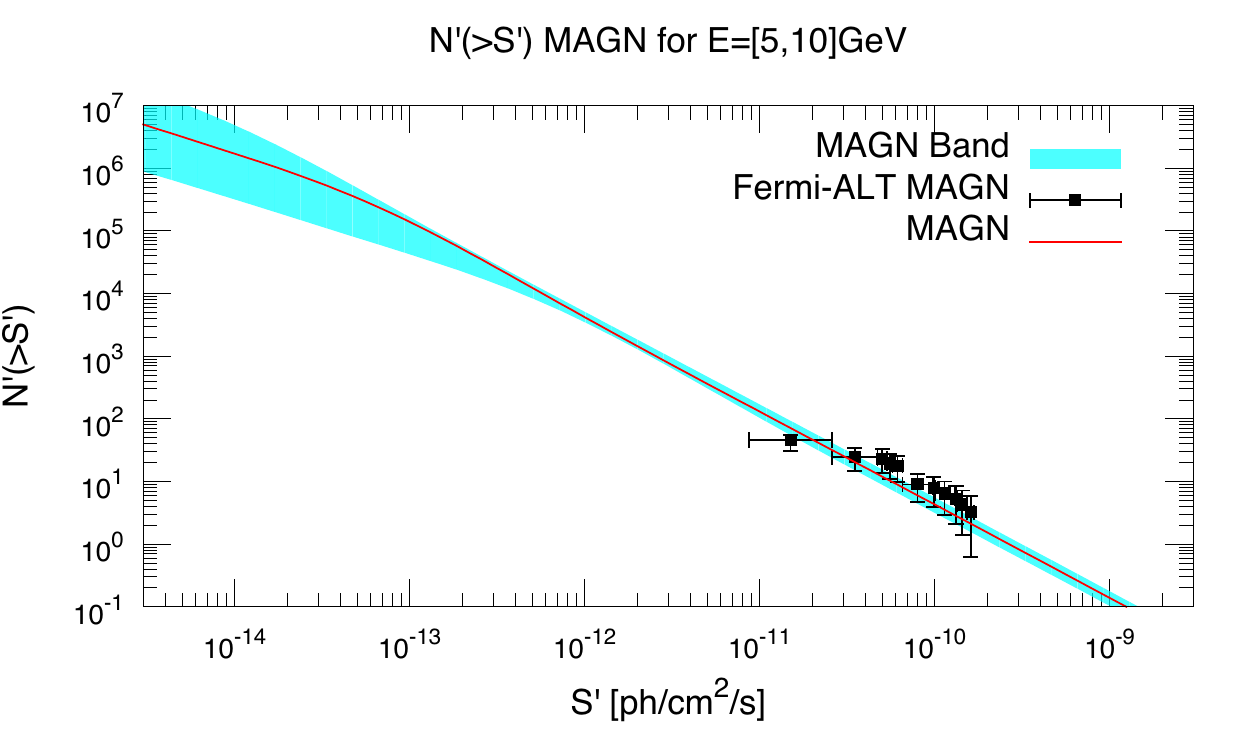}
\includegraphics[width=0.48\textwidth]{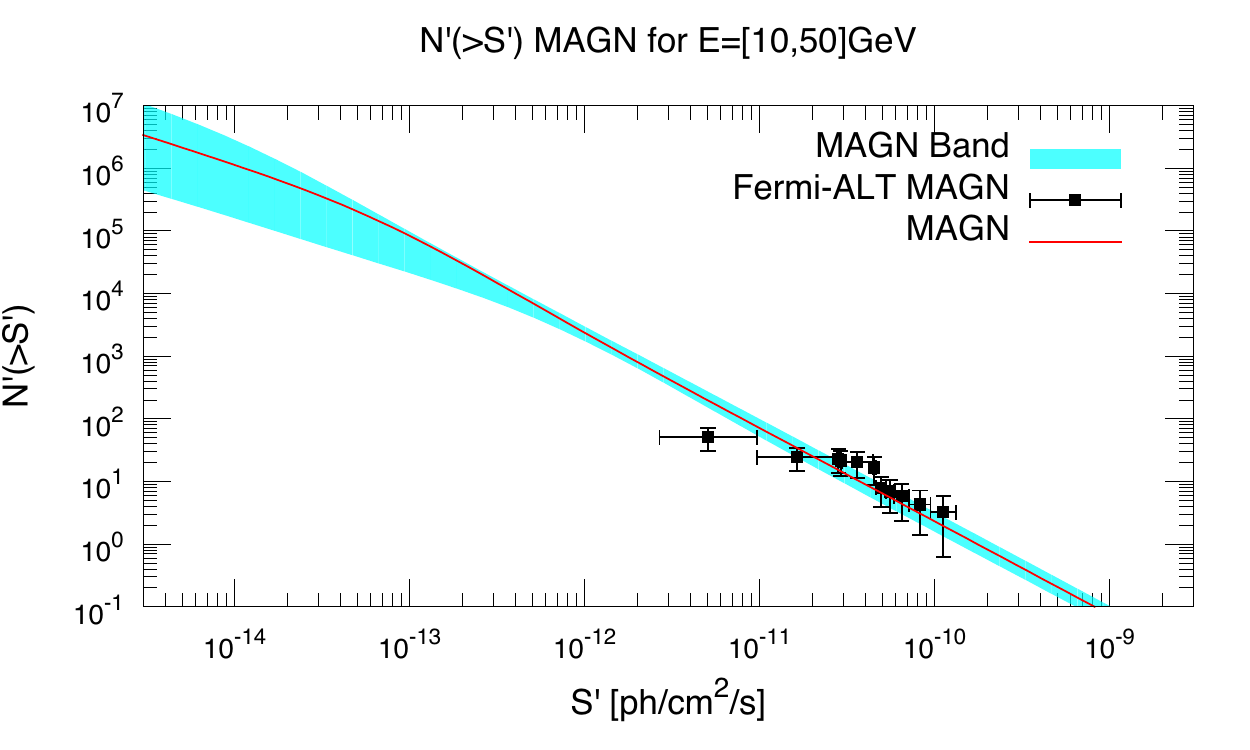}
\caption{As in Fig.~\ref{fig:ncounthsp}, but for MAGN.}
\label{fig:ncountmagn}
\medskip
\end{figure}

The resulting theoretical source count distributions $N'(>S')$ for the four $C_{\rm P}(E)$ energy bins used in the {\it Fermi}-LAT anisotropy measurement
are shown in Figs.~\ref{fig:ncounthsp} and \ref{fig:ncountmagn} for HSP BL Lacs and MAGN,  along with the  1$\sigma$ uncertainty band. 
We show also the experimental data points on $N'(>S')$, calculated with Eq.~\ref{eq:expNcount}
and using the relation Eq.~\ref{S'} between benchmark fluxes S and rescaled fluxes S': 
\begin{equation}
\label{eq:expNcount2}
   \displaystyle 	N'_{exp}(>S') = \sum^{N'_{S'}}_{i=1} \frac{1}{\omega(S_i(S'_i))}.
\end{equation}
The above procedure is only approximate, since in principle a new efficiency $\omega'(S')$ should be 
evaluated for the new energy band.  Alternatively, a proper conversion of $\omega$ between energy bands could be determined, however this
would require starting  from the full efficiency function $\omega(S,\Gamma)$ which is not available.
Nonetheless, it can be seen that the agreement between the data points and the 
theoretical predictions is reasonable.  The agreement can be seen as a cross-check
of the correctness of  the global rescaling procedure for the  $N(>S)$.
Note that for MAGNs the flux binning has been re-adjusted for each energy bin
due to the scarcity of sources available.
Note, further, that the source count distribution data points are used for illustrative purposes only in these figures, and are not used in any of the following calculations.

\subsection{$d^2N/(dSd\Gamma)$ energy range rescaling}
\label{sec:fit}
In this section we describe how we derive the rescaled double differential distribution $d^2N/(dS d\Gamma)$, 
which is the relevant quantity entering the calculation of the anisotropy term $C_{\rm P}(E)$. 
The quantity $d^2N/(dS d\Gamma)$ can be expressed in terms of the  $\gamma$-ray LF $\rho_{\gamma}$ as:
\begin{equation}
     \label{dNdGcounttwo}
	\frac{d^2N}{d\Gamma dS}(S,\Gamma) \approx \frac{\Delta \Omega}{\Delta S}  \int^{z_{\rm max}}_{z_{\rm min}} dz \int^{L_{\gamma}(S+\Delta S,\Gamma,z)}_{L_{\gamma}(S,\Gamma,z)} 
	dL_{\gamma} \frac{dV}{dz} \frac{dN}{d\Gamma} \rho_{\gamma}(L_{\gamma},z),
\end{equation} 
with $\Delta S$ sufficiently small.
Then, similarly to Eq.~\ref{N'count}, the rescaled $d^2N/(d\Gamma dS' )$ can be expressed as:     
\begin{equation}
     \label{dNdGcountrescal}
	\frac{d^2N}{d\Gamma dS'}(S',\Gamma) \approx  \frac{\Delta \Omega}{\Delta S'}  \int^{z_{\rm max}}_{z_{\rm min}} dz \int^{L_{\gamma}(S(S'+\Delta S',\Gamma),\Gamma,z)}_{L_{\gamma}(S(S',\Gamma),\Gamma,z)} 
	dL_{\gamma} \frac{dV}{dz} \frac{dN}{d\Gamma} \rho_{\gamma}(L_{\gamma},z),
\end{equation}     
again, with $\Delta S'$ sufficiently small.
We evaluate these expressions numerically, producing a table of $dN/(dS' d\Gamma)$ 
on a grid of $S'$ and $\Gamma$ values.
We also verified that choosing $\Delta S'$ and $\Delta S$ sufficiently small,
the result becomes independent of the actual chosen values.
Eq.~\ref{Cpdef} can then be used to calculate the anisotropy in each energy band.   

\section{Results}
\label{sec:results}
\begin{figure}
\centering
\includegraphics[width=0.80\textwidth]{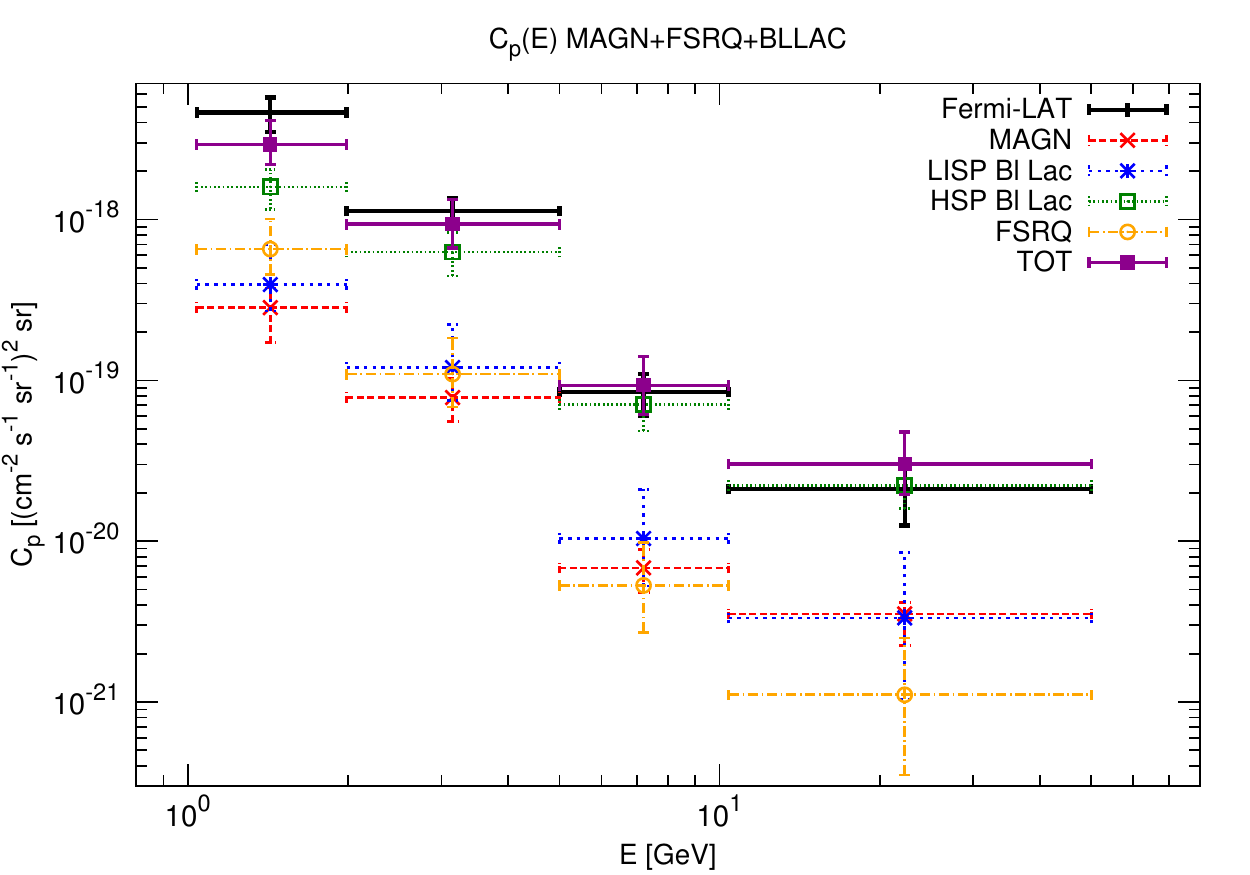}
\includegraphics[width=0.80\textwidth]{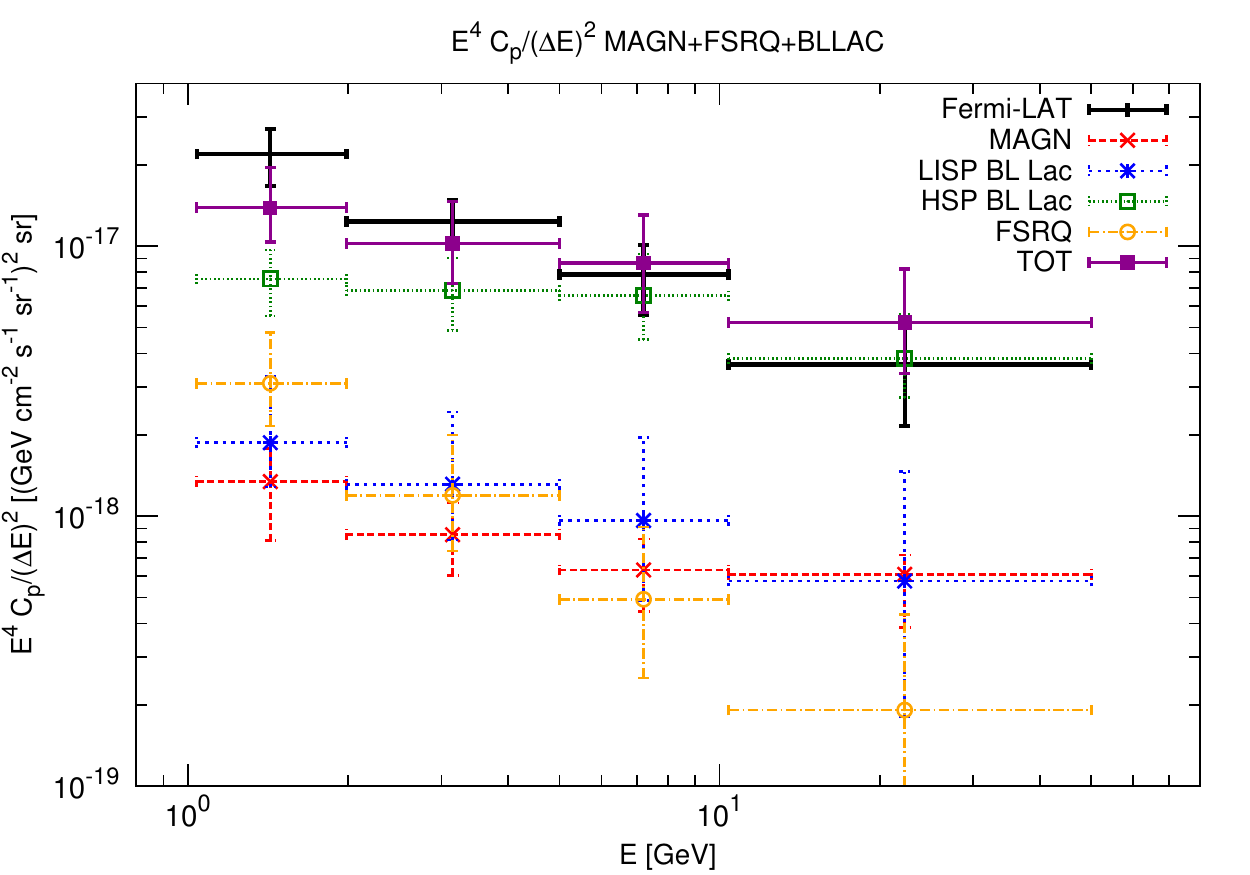}
\caption{The angular power $C_{\rm P}(E)$ for MAGN (red long-dashed points), LISP (blue short-dashed) and HSP BL Lacs (green dotted), FSRQs (yellow dot-dashed), and the total anisotropy (violet solid) from all the radio-loud AGN is shown in two different units ($C_{\rm P}(E)$ in the top panel and $E^4 C_{\rm P}(E)/(\Delta E)^2$ in the bottom panel). 
The data measured in the four energy bins analyzed by the {\it Fermi}-LAT Collaboration~\cite{2012PhRvD..85h3007A} are also shown (black solid points).}
\label{fig:Cpone}
\medskip
\end{figure}

The $C_{\rm P}(E)$ calculated using  the models and the method described in \S\ref{sec:N} and \S\ref{sec:fit} 
are shown in Fig.~\ref{fig:Cpone}.
We report results for the four energy bins used by the {\it Fermi}-LAT Collaboration in~\cite{2012PhRvD..85h3007A} 
(1.04-1.99 GeV, 1.99-5.00 GeV, 5.0-10.4 GeV and 10.4-50.0 GeV).  Results are shown for all the radio-loud AGN populations: FSRQs, BL Lacs, and MAGN, as well as their sum. 
For completeness the results are shown in two different ways, i.e., the anisotropy integrated
in each energy bin ($C_{\rm P}(E)$), and the quantity  $E^4 C_{\rm P}(E)/(\Delta E)^2$
which resembles a differential anisotropy spectrum.
The uncertainty for each theoretical bin has been derived from the uncertainty on the source count distribution 
given by the cyan band of the $N(>S)$ of Figs.~\ref{fig:ncounthsp} and~\ref{fig:ncountmagn}.
More precisely, the 1$\sigma$ band of the $N(>S)$ has first been transferred
to the $d^2N/(dS d\Gamma)$ distribution and then propagated to the $C_{\rm P}(E)$ through Eq.~\ref{Cpdef}.
In all the bins the population that gives the largest anisotropy is the HSP BL Lacs. 
Indeed, considering Figs.~\ref{fig:ncounthsp} and~\ref{fig:ncountmagn}, the HSP BL Lac population 
has about a factor of 3-5 times more sources in the bin 1.04-1.99~GeV with respect to MAGN at flux values just below the threshold of {\it Fermi}-LAT, 
which is the flux range where the unresolved sources contribute the most to the anisotropy. 
This factor of 3-5 between the number of HSP BL Lacs and MAGN translates
into the same factor for the angular power in the bin 1.04-1.99~GeV (see Fig.~\ref{fig:Cpone}).

\begin{figure}
\centering
\includegraphics[width=0.80\textwidth]{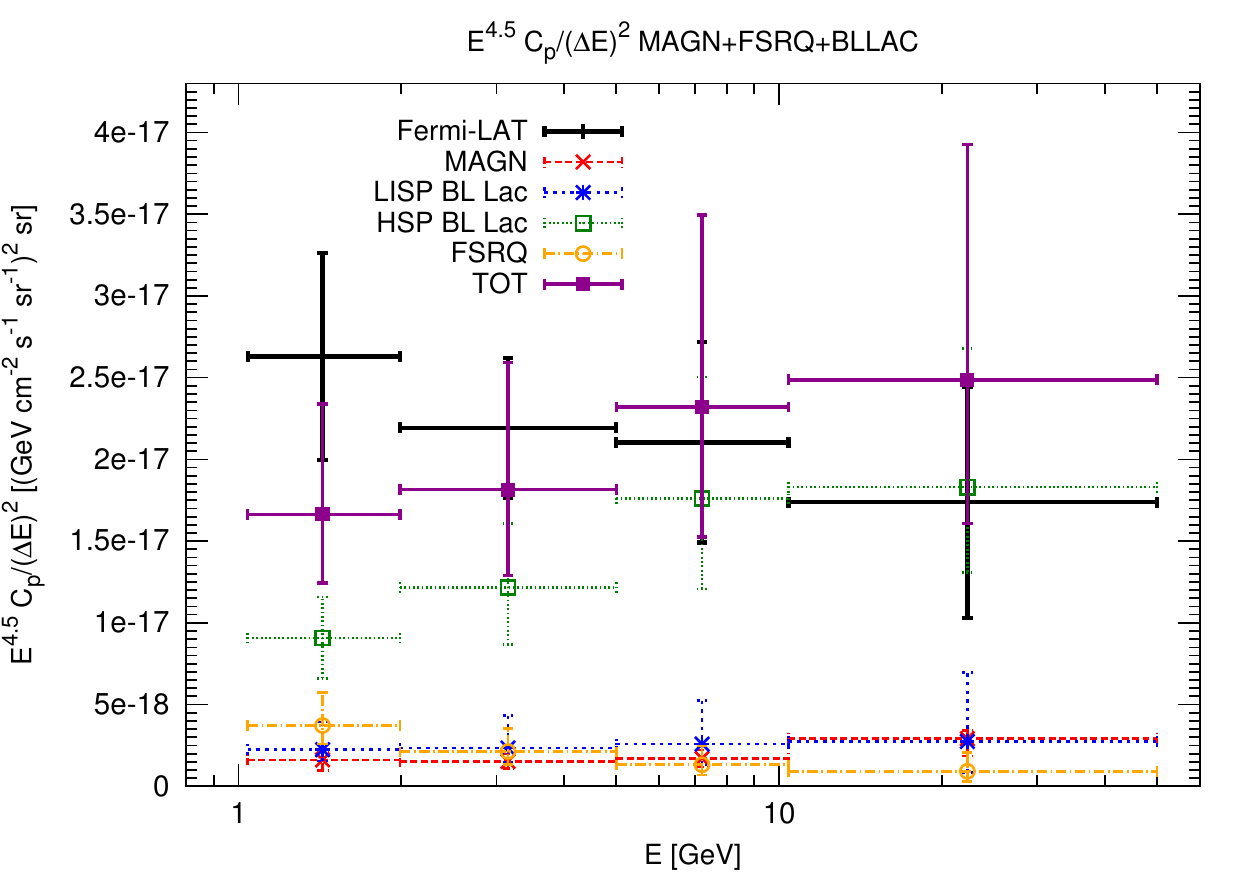}
\caption{The angular power $C_{\rm P}(E)$, in units of $E^{4.5} C_{\rm P}(E)/(\Delta E)^2$, for MAGN (red long-dashed points), LISP (blue short-dashed) and HSP BL Lacs (green dotted), FSRQs (yellow dot-dashed), and the total anisotropy (violet solid) from all the radio-loud AGN is shown. The data measured in the four energy bins analyzed by the {\it Fermi}-LAT Collaboration~\cite{2012PhRvD..85h3007A} are also shown (black solid points).}
\label{fig:Cplin}
\medskip
\end{figure}

We show in Fig.~\ref{fig:Cplin} the quantity $E^{4.5} C_{\rm P}(E)/(\Delta E)^2$ on a linear scale 
in order to illustrate more clearly  the differences between the data and the theoretical predictions. 
We see that radio-loud AGN can account for the total anisotropy
 measured in~\cite{2012PhRvD..85h3007A} by the {\it Fermi}-LAT Collaboration,
 the data and model predictions being compatible within the errors.
For example, in the energy range 1-10~GeV, the total theoretical expectation for the anisotropy from
radio-loud AGN is $C_{\rm P} = 9.3^{+3.5}_{-2.5} \cdot 10^{-18}$ $($cm$^{-2}$ s$^{-1}$ sr$^{-1})^2$ sr while 
the {\it Fermi}-LAT measurement is $(11.0\pm1.2) \cdot 10^{-18}$ $($cm$^{-2}$ s$^{-1}$ sr$^{-1})^2$ sr.
 However, a small trend (but within the errors) is visible, indicating
 that there could be room for further populations of sources contributing to the anisotropies 
 below 2 GeV, while above 10 GeV the anisotropy seems to be slightly overpredicted by the model.
 In this respect, a natural way to produce an anisotropy contribution only around a
 GeV would be to assume some contribution to the IGRB from unresolved 
 millisecond pulsars~\cite{SiegalGaskins:2010mp}. Nevertheless, in Ref. \cite{Calore:2014oga} it has been 
recently shown that   unresolved galactic millisecond pulsars contribute  by less than the percent level
to the measured anisotropy  in each energy range. 
We note also that we neglect the possible contribution from 
cascading GeV emission from hard-spectrum sources induced by propagation
of the TeV photons in the 
extra-galactic background light  (EBL) \cite{Aharonian:1993vz,2013MNRAS.432.3485V}.
This component is in principle sensitive to the presence
of inter-galactic magnetic fields \cite{2013MNRAS.432.3485V}. However,
the presence itself of the cascade emission
is still debated due to the possible
damping effect of plasma instabilities (see \cite{Chang:2013yia,Saveliev:2013jda,Sironi:2013qfa}).
Given the above uncertainties we will not investigate
further this component.

We check the robustness of the results by comparing the expected anisotropy associated with different models.
In particular the BL Lac models explored in~\cite{Ajello:2013lka},
like our benchmark BL Lac model of~\cite{DiMauro:2013zfa}, are not tuned 
explicitly to the anisotropy data and thus are suitable for a cross-check.
The models explored in~\cite{Ajello:2013lka} use  various different
$\rho_\gamma$  parametrized as PLE (pure luminosity evolution),
PDE (pure density evolution) or LDDE (luminosity dependent density evolution).
We adopt the PLE$_3$  model from Tab.~2 of~\cite{Ajello:2013lka}.
 
The model derived in~\cite{Harding:2012gk} adopts an LDDE $\gamma$-ray luminosity function 
based on the LF of observed X-ray AGNs, which is then rescaled to $\gamma$ rays
through a $L_\gamma$-$L_X$ relation.  The SED, instead, is based on a SED blazar sequence model
tuned to EGRET data. The free parameters of the model are then tuned
to the observed anisotropy~\cite{2012PhRvD..85h3007A}, resulting in a good match to the data themselves.
The predicted IGRB intensity above 1 GeV by this model is 
about 5\% which is similar (a bit lower) to the predictions of \cite{DiMauro:2013zfa}
and \cite{Ajello:2013lka} of about 15-20\%.

On the other hand~\cite{Chang:2013hia,Chang:2013yia} studied a new $\gamma$-ray 
propagation model associated with plasma instabilities of  ultra-relativistic $e^+e^-$ pairs.  
The instability then dissipates the kinetic energy of the TeV 
$e^+e^-$ pairs produced during the propagation of TeV $\gamma$ rays,
suppressing the development of the associated electro-magnetic cascade  
and heating the intergalactic medium. 
The effectiveness of the plasma instability mechanism is, nonetheless, still being
investigated (see, e.g., \cite{Saveliev:2013jda,Sironi:2013qfa}).
The blazar LF used in \cite{Chang:2013yia} is based on the LF of optical and X-ray observed quasars,
rescaled to $\gamma$-ray energies while the blazar SED is modeled as a broken power-law.
No parameter of the model is tuned to $\gamma$-ray data.
This \emph{a priori} model matches well the intensity and spectrum of the IGRB above few GeVs,
thus being compatible with 100\% of the IGRB contrary to the models described above
which  associate to blazars only a small fraction of the IGRB.
The model, indeed, largely over-predicts the anisotropy. The authors claim, however,
that the IGRB anisotropy measured in \cite{2012PhRvD..85h3007A} has been 
substantially under-estimated \cite{2013arXiv1308.0015B}.

Finally, the model developed in~\cite{2011ApJ...736...40S} is 
difficult to use for comparison in the present study since the full LF is not available. Nonetheless, 
the model has already been shown to be significantly in tension with the anisotropy data~\cite{Cuoco:2012yf}.

\begin{figure}
\centering
\includegraphics[width=0.70\textwidth]{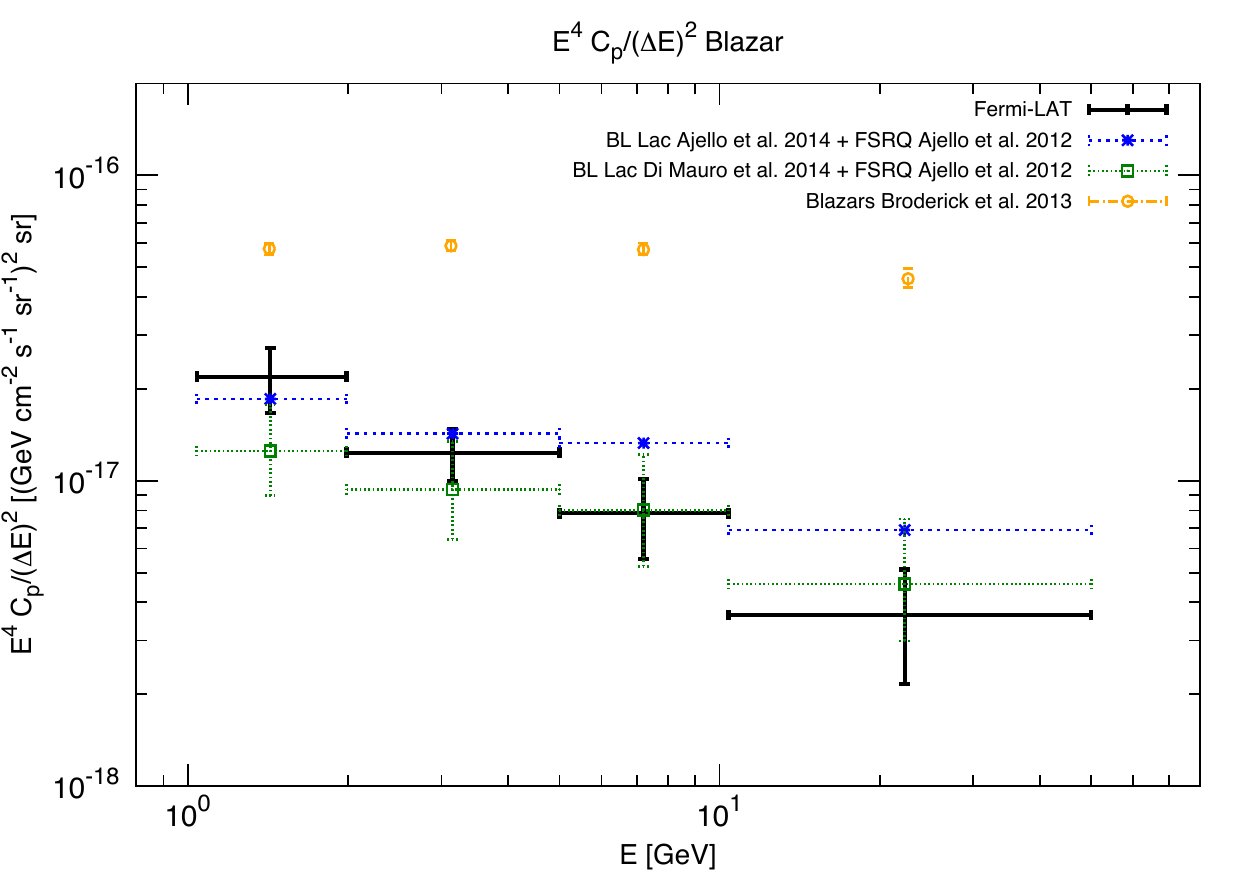}
\caption{The angular power $C_{\rm P}(E)$ for different models of blazars is shown.
Blue short-dashed: FSRQs from~\cite{Ajello:2011zi} and BL Lacs from~\cite{Ajello:2013lka}.
Green dotted: BL Lacs from~\cite{DiMauro:2013zfa} and FSRQs from~\cite{Ajello:2011zi}.
Orange circles: blazar anisotropies from \cite{Chang:2013yia}. 
The data measured in~\cite{Abdo:2010nz} by the {\it Fermi}-LAT Collaboration (black solid points) are also shown.}
\label{fig:Cpblazars}
\medskip
\end{figure}

We show in Fig.~\ref{fig:Cpblazars} the anisotropy results for our benchmark case
(FSRQs from~\cite{Ajello:2011zi} and BL Lacs from~\cite{DiMauro:2013zfa})
compared with the case of FSRQs from~\cite{Ajello:2011zi} and BL Lacs from~\cite{Ajello:2013lka},
and the blazar model of \cite{Chang:2013yia},
and with the observed anisotropy.
We do not show the model anisotropies from \cite{Harding:2012gk}
since after the fit they closely match the data.
The model of~\cite{Ajello:2013lka} yields a larger anisotropy than the model of~\cite{DiMauro:2013zfa} in all energy bands, although still compatible with the measured {\it Fermi}-LAT anisotropy.
Note that for the model of~\cite{Ajello:2013lka} we report the predicted anisotropy
without uncertainties, adopting the central value for each of the parameters given
in Tab.~2 of~\cite{Ajello:2013lka} for the PLE$_3$ model without considering their errors.
The parameters are in fact strongly correlated and propagating their uncertainty
to the $C_{\rm P}$ would require knowledge of the full covariance matrix,
which is not available.
The differences between the $C_{\rm P}(E)$ in Fig.~\ref{fig:Cpblazars} are due to the differences in the $\gamma$-ray emission models and the catalogs used to calculate the average source parameters. In particular, the models for FSRQs and BL Lacs in~\cite{Ajello:2011zi,Ajello:2013lka} use the 1FGL catalog~\cite{Abdo:2010ru} and use both an LDDE and PLE $\gamma$-ray luminosity function. The FSRQ SED is calibrated by combining {\it Fermi}-LAT data with X-ray measurements from the {\it Swift} Burst Alert Telescope (BAT)~\cite{2004ApJ...611.1005G} while the BL Lac $\gamma$-ray SED is modeled using a simple power-law.
The BL Lac model derived in~\cite{DiMauro:2013zfa} is calibrated using the 2FGL {\it Fermi}-LAT catalog~\cite{Fermi-LAT:2011iqa} and uses both an LDDE and PLE $\gamma$-ray luminosity function. 
The $\gamma$-ray SED is studied by adding to the 2FGL catalog also the first {\it Fermi}-LAT catalog of Sources Above 10 GeV (1FHL)~\cite{TheFermi-LAT:2013xza} and the TeV measurements from available IACTs~\cite{tevcat}.

\begin{figure}
\centering
\includegraphics[width=0.80\textwidth]{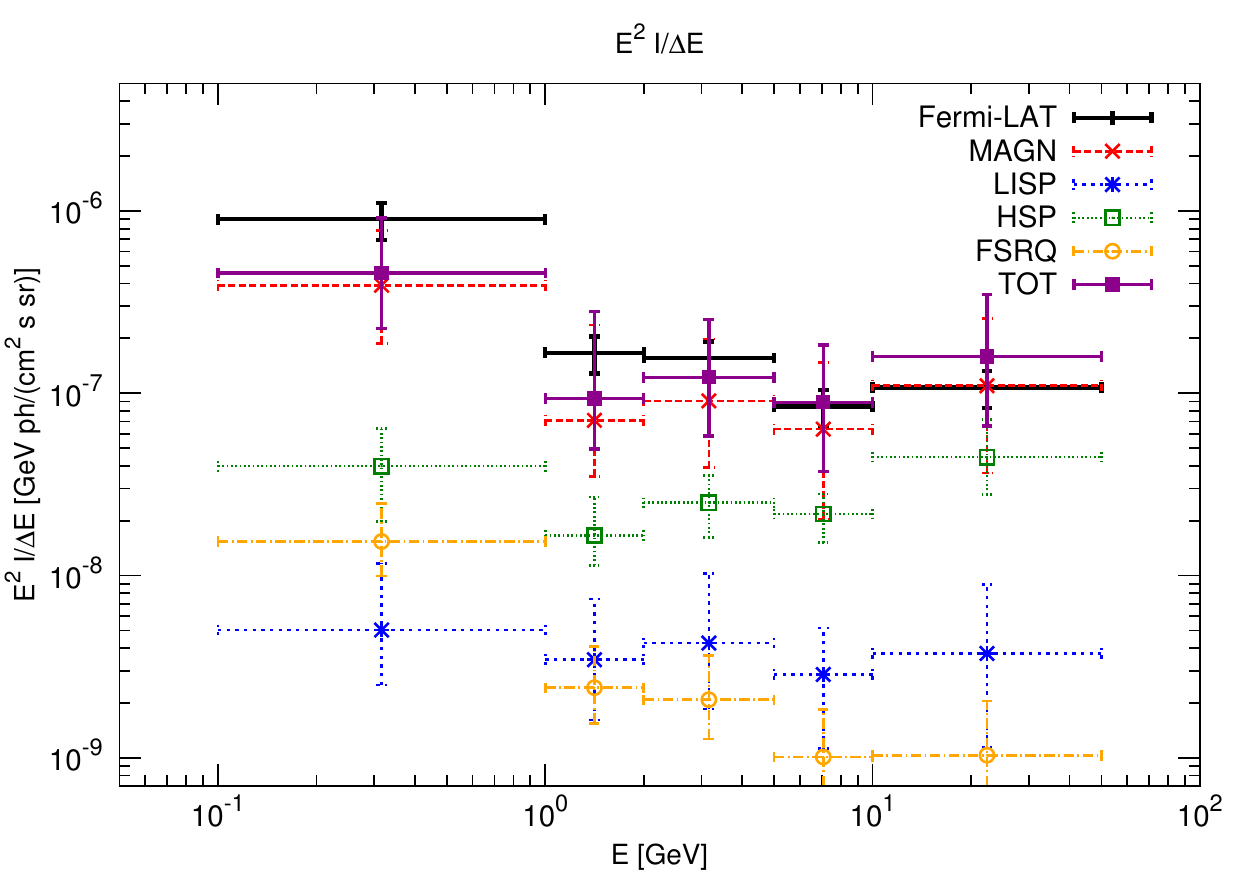}
\caption{The integrated flux $E^2 I / \Delta E$ for MAGN (red long-dashed points), LISP (blue short-dashed) and HSP BL Lacs (green dotted), FSRQs (yellow dot-dashed) and the total flux (violet solid) from all the radio-loud AGN are shown. Here we also show the data measured in~\cite{Abdo:2010nz} by the {\it Fermi}-LAT Collaboration 
rebinned into the 5 shown energy bins (black solid points).}
\label{fig:Ione}
\medskip
\end{figure}

Using the same formalism we can also calculate the expected
flux contribution from radio-loud AGN to the IGRB~\cite{Abdo:2010nz}.
The integrated $\gamma$-ray flux $I$ in a given energy bin can be written as: 
\begin{equation}
     \label{Idef}
        I(E_0 \leq E \leq E_1) = \int_{\Gamma_{\rm min}}^{\Gamma_{\rm max}} d\Gamma \int^{S_t(\Gamma)}_0 S \frac{d^2N}{dS d\Gamma} dS.
    \end{equation} 
The contribution to the IGRB intensity from MAGN, FSRQs and BL Lacs  are
shown in Fig.~\ref{fig:Ione} with the same notation and in the same energy bins as Fig.~\ref{fig:Cpone}.
For a visual comparison with the anisotropy results we rebin the Fermi-LAT
data of \cite{Abdo:2010nz} into 5 new energy bins, 4 of which match the $C_P$ binning. 
The rebinned data are used only for visual purposes in Fig.~\ref{fig:Ione}.
We emphasize that the radio-loud AGN emission can explain both the angular power and the intensity flux measured by {\it Fermi}-LAT\@, given that these two observables trace different features of 
the $\gamma$-ray population. Indeed the population that gives the largest contribution to anisotropy is, as we have already stressed, HSP BL Lacs because the unresolved part near the flux 
threshold of the {\it Fermi}-LAT is more populated. On the other hand, concerning the IGRB intensity, MAGN sources give the largest $\gamma$-ray flux, as already found in~\cite{DiMauro:2013zfa}, 
because at fluxes much below the {\it Fermi}-LAT sensitivity, the MAGN population is expected to have many more sources with respect to the other radio-loud AGN populations.
Moreover FSRQ and LISP BL Lac sources provide only a small fraction of both the anisotropy and flux of the IGRB at energies larger than a few tens of GeV since their emission is suppressed by the 
cut-off or by the steep spectrum of their SED.

Anisotropy and flux are two complementary observables that can be used together 
to set constraints on the radio-loud AGN $\gamma$-ray emission.
In Tab.~\ref{tab:igrbul} and  \ref{tab:anisul} we display the best-fit and  $\pm1\sigma$ percentage 
values of the contribution of each unresolved population to the measured IGRB anisotropy and intensity~\cite{2012PhRvD..85h3007A,Abdo:2010nz}.
These percentage values are derived with respect to the central values of the data.
As already pointed out above, radio-loud AGN sources can in principle entirely explain both the anisotropy and intensity data within the theoretical and experimental 1$\sigma$ uncertainties.
However the uncertainty associated with the diffuse $\gamma$-ray emission from MAGN is large, due to  the small sample of MAGN present in the {\it Fermi}-LAT 2FGL 
catalog~\cite{Fermi-LAT:2011iqa}. The $\gamma$-ray flux from these objects could be more severely constrained by future {\it Fermi}-LAT catalogs with a larger number of sources and with a deeper 
knowledge of the  correlation between the $\gamma$-ray and radio emission from the center of those sources.

LISP BL Lacs and FSRQs have a subdominant contribution to both the IGRB intensity and anisotropy, while MAGN give the largest contribution to the IGRB intensity and HSP BL Lacs the largest 
contribution to the anisotropy.
The radio-loud AGN population gives the smallest fractional contribution to $C_{\rm P}(E)$ and the measured intensity in the lowest energy bins, 
namely at 1-2~GeV for the anisotropy and 0.1-1~GeV for the intensity, 
where the central values of the theoretical predictions are approximately half of the central value of the data.
At these energies other populations could give a sizable contribution to the anisotropy and the diffuse emission. Star-forming galaxies have been widely studied and are expected to contribute in the 
range 0.1-10 GeV, see~\cite{Ackermann:2012vca} and references therein.
Millisecond pulsars, the most numerous Galactic population,
have been shown to give negligible contribution to both the IGRB intensity and anisotropy \cite{Calore:2014oga}. 

\begin{table}
\begin{center}
\begin{tabular}  {||c|c|c|c|c|c|c|c||}
\hline
Energy (GeV)	&	LISP BL Lac	&	HSP BL Lac	&	BL Lac	&	FSRQ	&	Blazars	&	MAGN	&	AGN	\\
\hline
\hline
  1.04-1.99	&   $8.5^{+6.5}_{-2.7}$	&	$34.5^{+9.5}_{-9.4}$		&  $43^{+16}_{-12}$		&	$14.1^{+7.7}_{-4.3}$	&	$57^{+24}_{-16}$ &	$6.1^{+2.0}_{-2.4}$	&	$63^{+26}_{-19}$	\\
  1.99-5.00	&   $10.6^{+9.1}_{-4.0}$	&	$55.5^{+17.7}_{-16.1}$	&	$66^{+27}_{-20}$	&	$9.7^{+6.5}_{-3.7}$	&	$76^{+33}_{-24}$ &	$6.9^{+2.2}_{-2.0}$	&	$83^{+35}_{-26}$	\\
 5.0-10.4	&   $12.3^{+12.6}_{-6.1}$	&	$83.7^{+35.4}_{-26.2}$	&	$96^{+48}_{-32}$	&	$6.3^{+5.4}_{-3.1}$	&	$102^{+53}_{-35}$ &	$8.1^{+2.4}_{-2.4}$  &	$110^{+56}_{-38}$	\\
 10.4-50.0	&   $15.7^{+24.3}_{-10.8}$	&	$105^{+49}_{-30}$		&	$121^{+73}_{-41}$	&	$5.3^{+6.6}_{-3.5}$	&	$126^{+80}_{-44}$ &	$16.7^{+3.0}_{-6.1}$	&	$143^{+82}_{-50}$	\\
\hline
\hline
\end{tabular}
\caption{
 Model predictions, together with 1$\sigma$ uncertainties,  for the contribution to the IGRB anisotropy from the unresolved LISP BL Lacs, HSP BL Lacs, total from BL Lacs, FSRQs, all blazars (BL Lacs 
 + FSRQs), MAGN, and all radio-loud AGN (blazars + MAGN) in the four energy bins indicated in the first column. 
  The predictions are given in terms of the percentage of the central value of the experimental measure.}
\label{tab:igrbul}
\end{center}
\end{table}
\begin{table}
\begin{center}
\begin{tabular}  {||c|c|c|c|c|c|c|c||}
\hline
Energy (GeV)	&	LISP BL Lac	&	HSP BL Lac	&	BL Lac	&	FSRQ	&	Blazars	&	MAGN	&	AGN	\\
\hline
\hline
  0.1-1	&   $0.56^{+0.73}_{-0.28}$	&	$4.4^{+2.7}_{-2.2}$	&	$5.0^{+3.4}_{-2.5}$	&	$1.7^{+1.0}_{-0.6}$	&	$6.7^{+4.5}_{-3.1}$ &	$43^{+43}_{-23}$	&	$50^{+48}_{-26}$	\\
  1.04-1.99	&   $2.1^{+2.4}_{-1.1}$	&	$10.0^{+6.2}_{-3.1}$	&	$12^{+9}_{-4}$		&	$1.5^{+1.0}_{-0.5}$	&	$14^{+10}_{-5}$ &	$43^{+100}_{-22}$	&	$56^{+109}_{-26}$	\\
 1.99-5.00	&   $2.7^{+3.8}_{-1.5}$	&	$16.0^{+6.6}_{-5.7}$	&	$19^{+10}_{-7}$		&	$1.3^{+1.0}_{-0.5}$	&	$20^{+11}_{-8}$ &	$58^{+69}_{-33}$	&	$78^{+80}_{-41}$	\\
 5.0-10.4	&   $3.4^{+2.7}_{-2.1}$	&	$26^{+8}_{-7}$		&	$29^{+10}_{-10}$	&	$1.2^{+1.0}_{-0.5}$	&	$30^{+11}_{-10}$ &	$75^{+100}_{-51}$	&	$105^{+110}_{-61}$	\\
10.4-50.0	&   $3.5^{+4.9}_{-2.4}$	&	$41^{+25}_{-16}$	&	$45^{+30}_{-18}$	&	$0.96^{+0.94}_{-0.53}$	&	$46^{+31}_{-19}$ &	$102^{+136}_{-68}$	&	$148^{+167}_{-87}$	\\
\hline
\hline
\end{tabular}
\caption{As in Table \ref{tab:igrbul}, but for the IGRB intensity.}
\label{tab:anisul}
\end{center}
\end{table}

\begin{figure}
\centering
\includegraphics[width=0.80\textwidth]{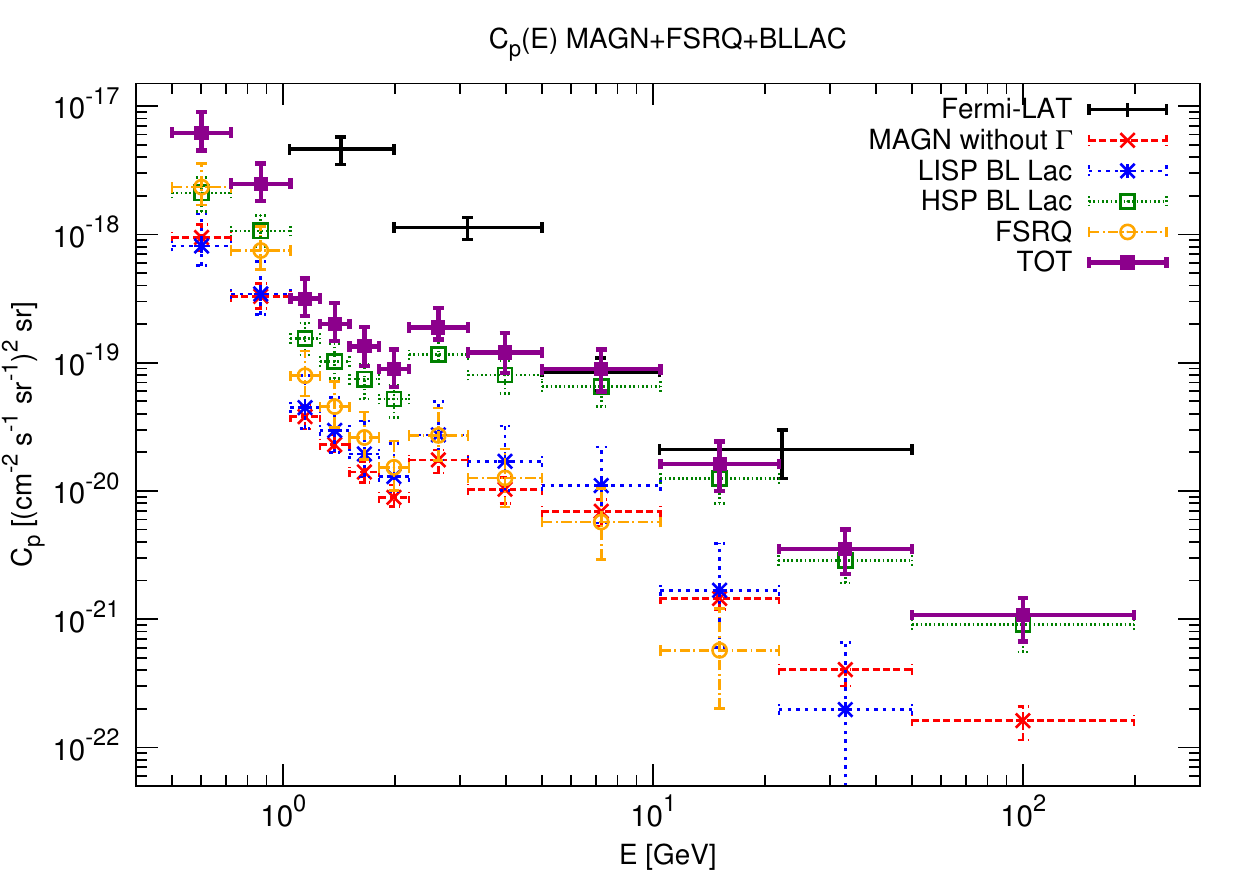}
\includegraphics[width=0.80\textwidth]{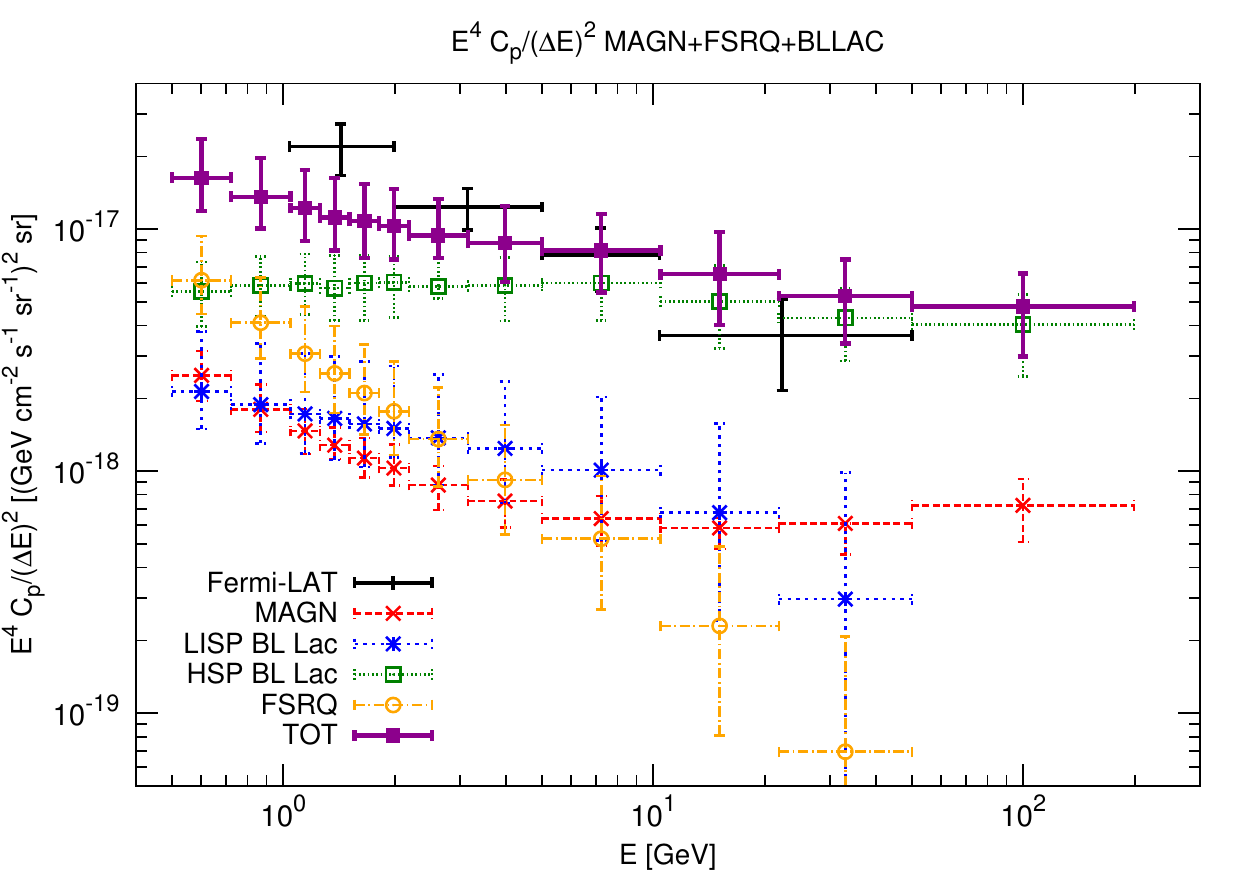}
\caption{In this figure we display the angular power $C_{\rm P}(E)$, in two different units ($C_{\rm P}(E)$, top, and $E^4 C_{\rm P}(E)/(\Delta E)^2$, bottom), 
for MAGN (red long-dashed), LISP (blue short-
dashed) and HSP BL Lac (green dotted), FSRQ (yellow dot-dashed) and the total anisotropy (violet solid) in 12 energy bins, from 0.5 to 200 GeV. For illustrative purposes we also show the {\it Fermi}-
LAT data~\cite{2012PhRvD..85h3007A} (grey solid points).  Note that in the top panel the quantity shown is integrated within each energy bin, and thus its value depends on the size of 
the bin considered; the binning differs between the measured data points and the predictions.}
\label{fig:Cptwo}
\medskip
\end{figure}

Finally, one has to note that  with increasing statistics  a new measurement of the anisotropy
will likely be performed in smaller energy bins and in an extended energy range.
For reference, we thus provide predictions in the range 500~MeV -- 200~GeV in 12 energy bins.
Results are shown in Fig.~\ref{fig:Cptwo},  where the energy dependence of
the different populations is more clearly apparent.
Again, as for Fig.~\ref{fig:Cpone}, we show predictions in two complementary ways.
The top panel shows the total anisotropy integrated
in each energy bin ($C_{\rm P}(E)$), while the bottom panel shows  $E^4 C_{\rm P}(E)/(\Delta E)^2$.
It can be seen that at high energies, above about 10 GeV, only MAGN and HSP BL Lac
 sources still give a sizable contribution, due to the FSRQ and LISP BL Lac SED cut-off above these energies.
The dip-like feature at about 1 GeV in the top panel is
due to the fine subdivision of the 1-2 GeV range into 4 bins,
consequently partitioning the total anisotropy in 4 bins.
The effect disappears in the bottom panel differential
spectrum where each bin is weighted by its energy width.

\section{Predicted anisotropy for CTA}\label{sec:cta}

In this section we derive a forecast for the anisotropy
from  unresolved sources expected in  CTA observations.
To derive these predictions we use the same models and formalism considered in the previous sections
extrapolated in the TeV energy range. 
CTA~\cite{Consortium:2010bc,Dubus:2012hm} is a project in the development phase
 which will study  very high-energy $\gamma$-rays from a few tens of GeV up to possibly 100 TeV.
We use  as benchmarks four different values of the CTA flux threshold (see Eq.~\ref{Cpdef}) 
for point source detection, in units of the Crab flux, namely 5, 10, 50 and 100 mCrab. 
To define the Crab flux we assume a power-law spectra with a photon index of 2.63 
normalized to a flux of $3.45 \cdot 10^{-11}$ cm$^{-2}$ s$^{-1}$ TeV$^{-1}$ at 1 TeV
as measured by~\cite{Aharonian:2006pe}.
We consider predictions in the energy range $[0.1,10]$ TeV, divided 
into four energy bins (0.1-0.3, 0.3-1, 1-3 and 3-10~TeV) in which we have computed 
the unresolved angular power for radio-loud AGN varying the flux threshold in the range [5,100] mCrab flux.
For simplicity, we do not consider any index bias, but we assume a single threshold in each energy bin
independent from the spectral index $\Gamma$.
We expect the anisotropy variation due to different thresholds to be
much larger than spectral index bias, which,
in any case, is difficult to model at present without precise knowledge 
of CTA's instrumental response functions.

\begin{figure}
\centering
\includegraphics[width=0.47\textwidth]{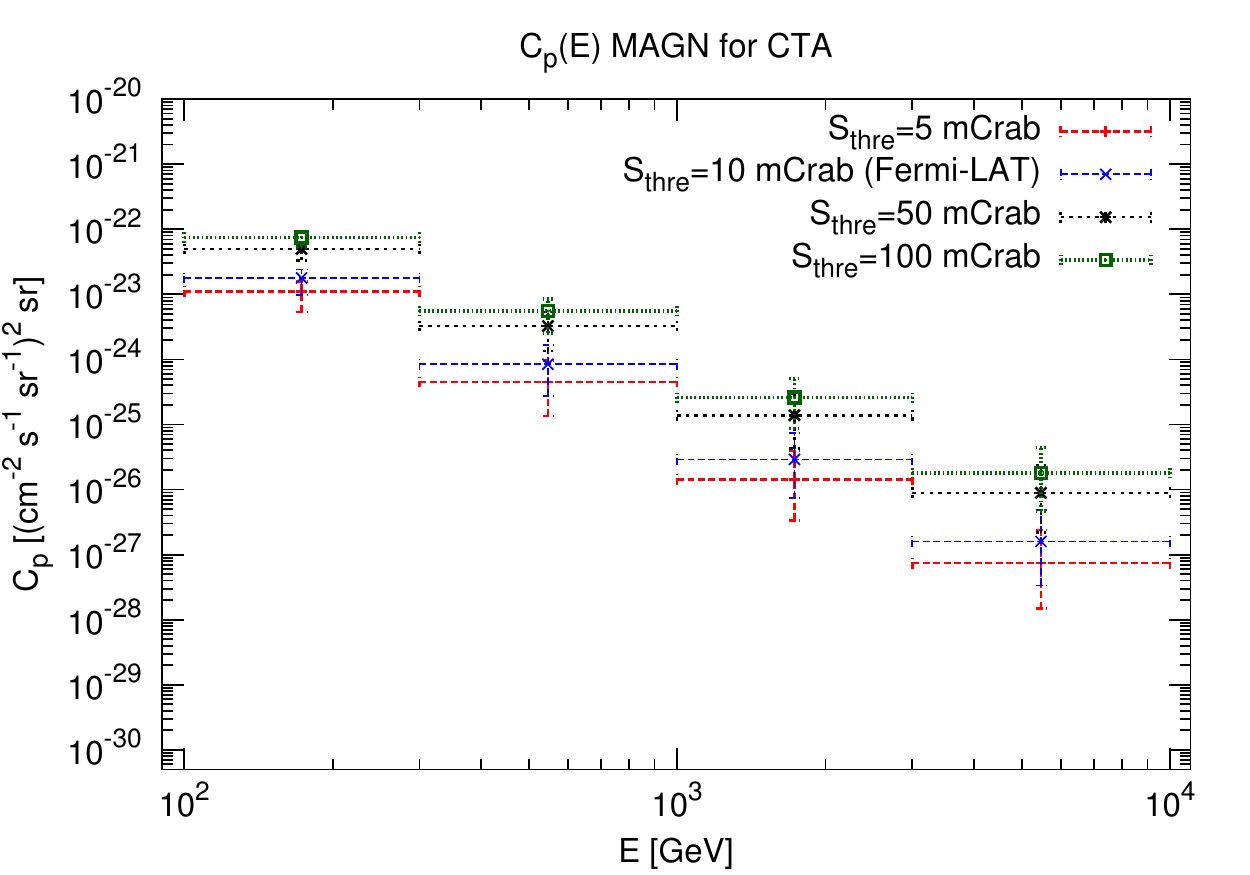}
\includegraphics[width=0.47\textwidth]{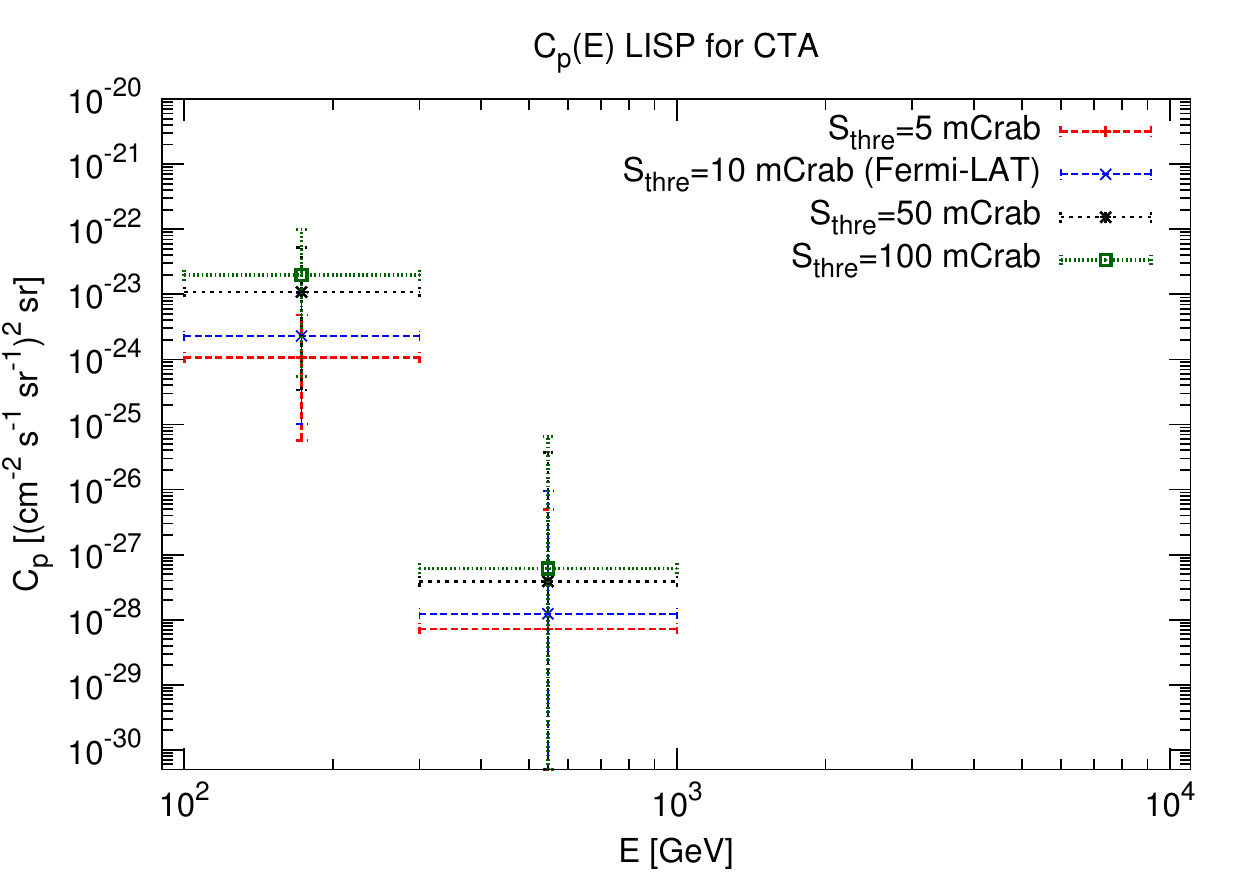}
\includegraphics[width=0.47\textwidth]{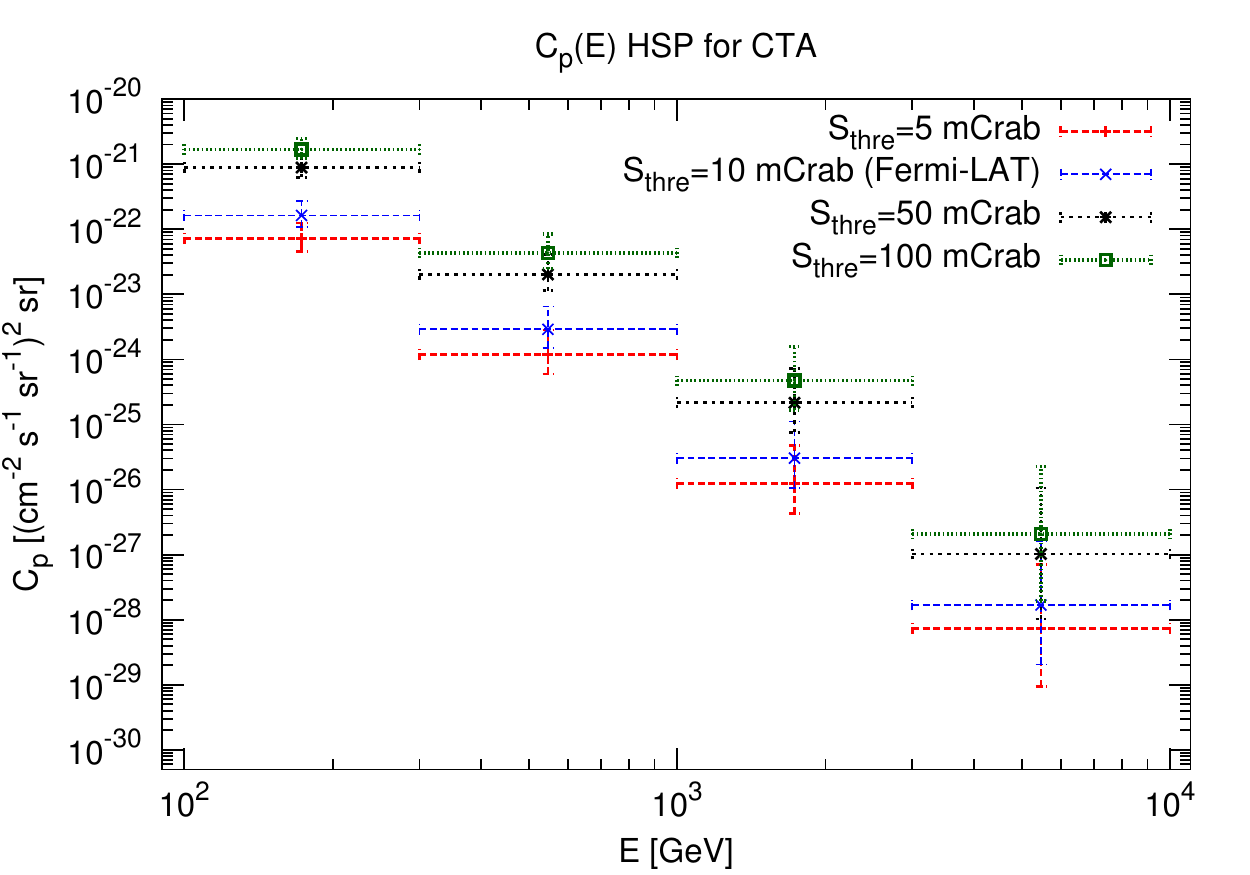}
\includegraphics[width=0.47\textwidth]{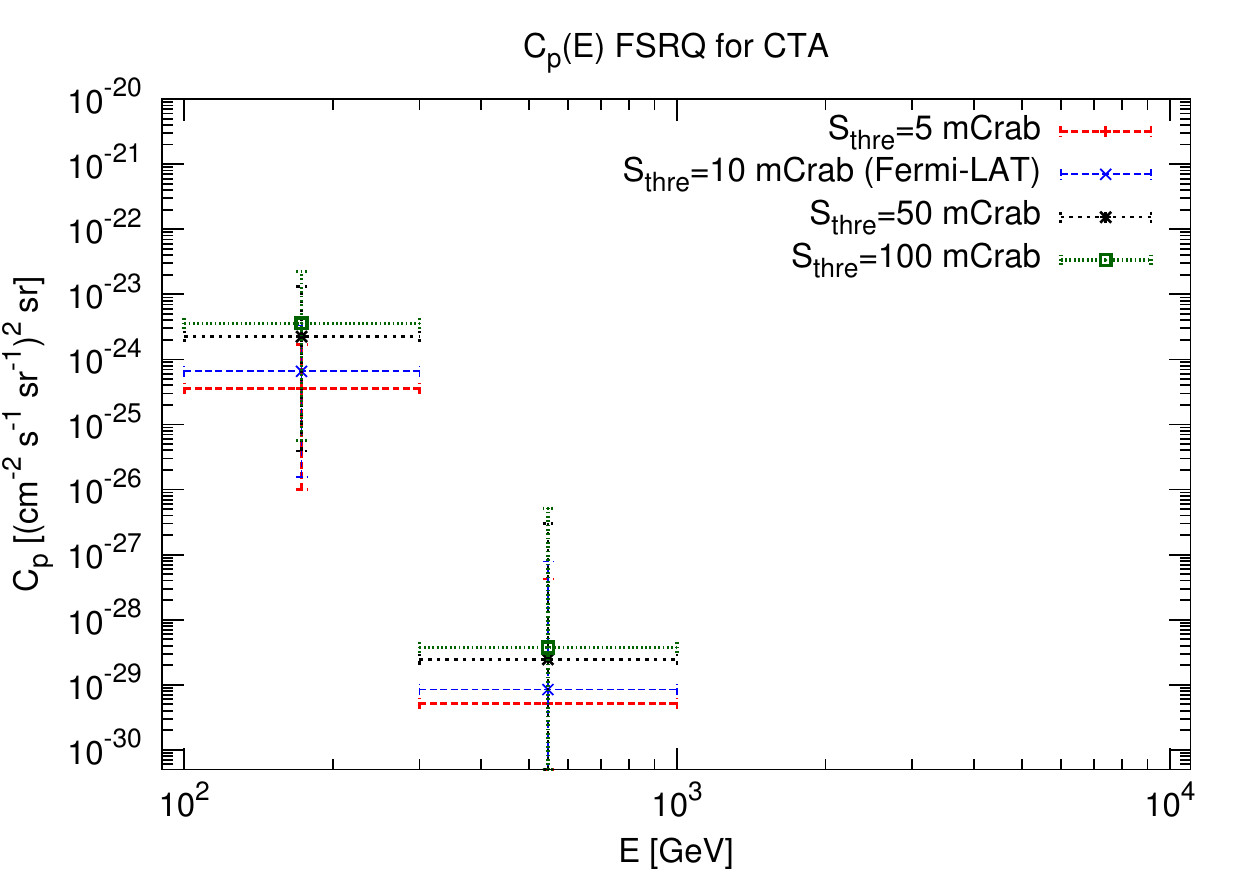}
\caption{Predicted angular power $C_{\rm P}(E)$ for different expected CTA sensitivities (in units of the Crab flux), in 4 energy bins from 100 GeV to 10 TeV.  The top left and top right panels refer to unresolved MAGN and LISPs, respectively, while the bottom left and bottom right panels refer to HSP and FSRQ sources, respectively.}
\label{fig:aniscta}
\medskip
\end{figure}

\begin{figure}
\centering
\includegraphics[width=0.70\textwidth]{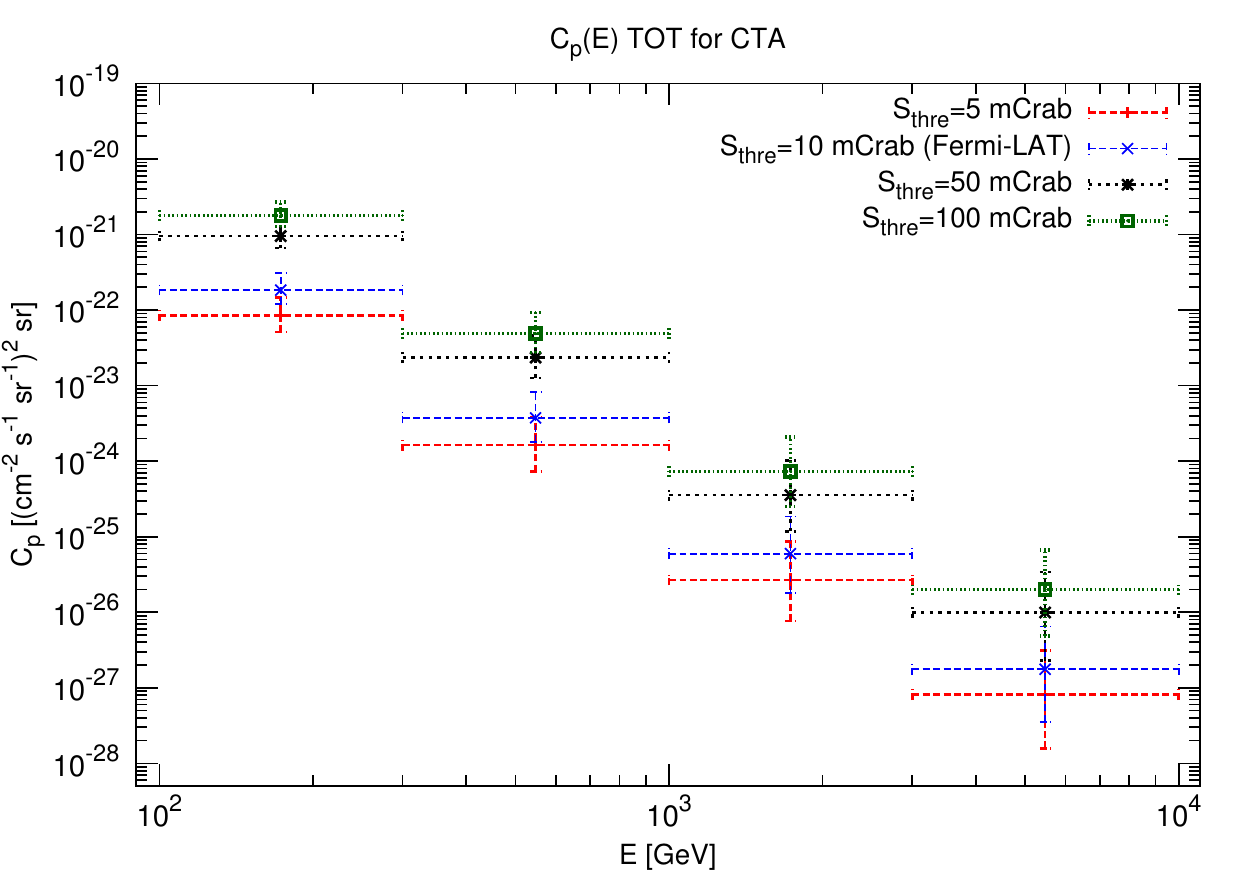}
\caption{Total angular power $C_{\rm P} (E)$ from all the unresolved radio-loud AGN (sum of the contributions shown in Fig.~\ref{fig:aniscta}) in the 4 energy bins listed in the text and for different
threshold fluxes in units of the Crab flux.}
\label{fig:anisctatot}
\medskip
\end{figure}

The anisotropy predictions are shown in Fig.~\ref{fig:aniscta} 
for the separated MAGN, LISP and HSP BL Lac, and FSRQ populations
while their sum is shown in Fig.~\ref{fig:anisctatot}.
As already noticed for Fig.~\ref{fig:Cptwo}, above 100 GeV the anisotropy is
dominated by the HSP and MAGN populations while 
LISP and FSRQs provide a sub-dominant contribution. 
The CTA experiment could perform observations both in single-source pointing mode and in survey mode.
With the single-source configuration the flux sensitivity is a few mCrab, while in the survey strategy it could be a few tens of mCrab~\cite{Dubus:2012hm}. Hence, in the first case the unresolved 
angular power for $E>100$ GeV could be of the order of $10^{-22}$ (cm$^{-2}$ s$^{-1}$ sr$^{-1}$)$^2$ sr.
On the other hand, with survey mode observations the unresolved angular power for $E>100$ GeV could be in the range $[10^{-22},10^{-21}]$ (cm$^{-2}$ s$^{-1}$ sr$^{-1}$)$^2$ sr.
 
A way to make these predictions more intuitive is to compare CTA's
sensitivity with that of the {\it Fermi}-LAT, at least in the energy range 10-1000 GeV where there is
a partial overlap between the two instruments' operating energies.
Looking at the sensitivity map in the First High Energy {\it Fermi}- Catalog (1FHL)~\cite{TheFermi-LAT:2013xza}, 
the flux sensitivity at high latitudes ($b>10$) is about $S_f = 9 \cdot 10^{-11}$ ph/cm$^2$/s for energies larger than 10 GeV.
On the other hand, the Crab flux above 10 GeV for the power-law model assumed above is $8.4 \cdot 10^{-9}$ ph/cm$^2$/s
which translates to a {\it Fermi}-LAT sensitivity of about 10 mCrab.
For comparison, we thus indicate in Figs.~\ref{fig:aniscta} and \ref{fig:anisctatot}  the 10 mCrab
sensitivity as ``{\it Fermi}-LAT-equivalent sensitivity''.

\begin{figure}
\centering
\includegraphics[width=0.70\textwidth]{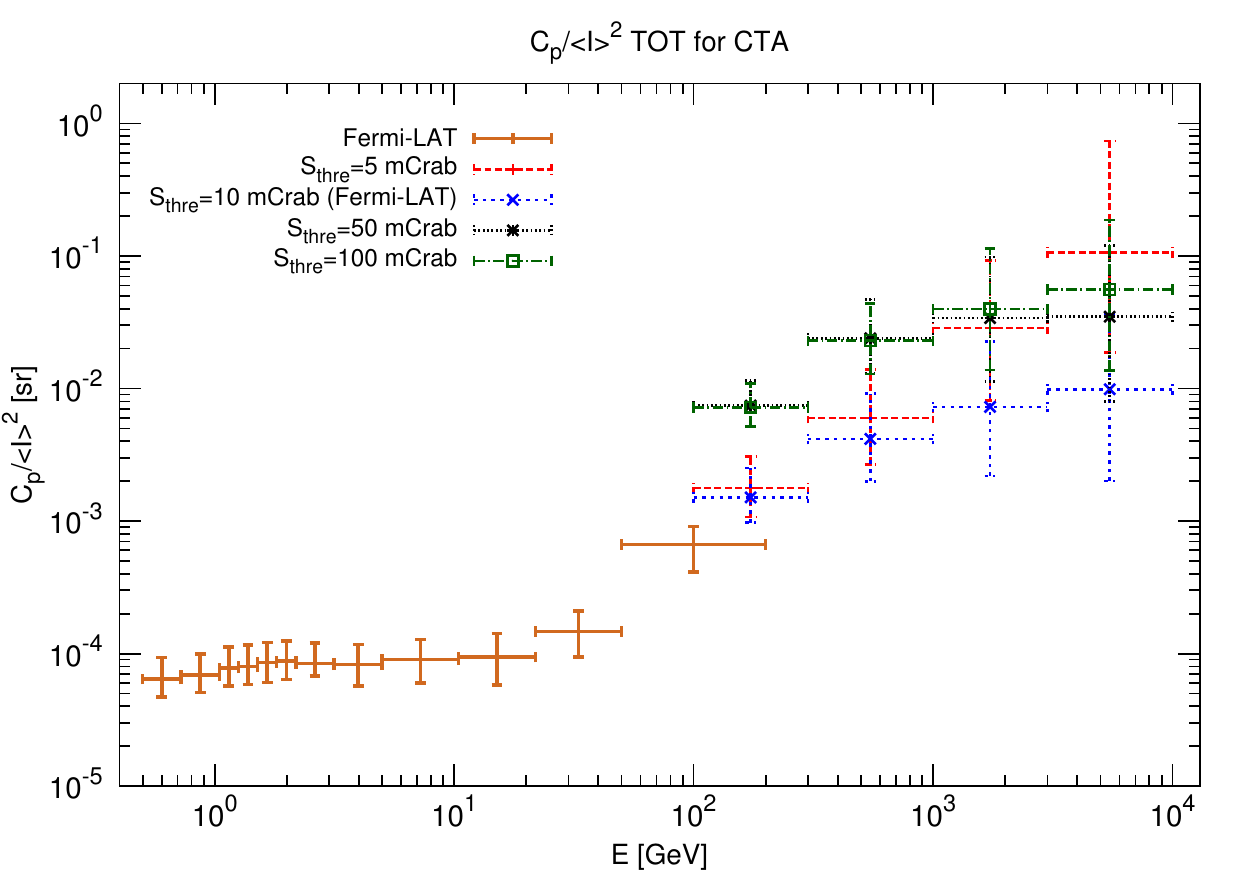}
\caption{Total angular power (sum from all radio-loud AGN populations) divided by the square of the intensity $C_{\rm P}/\langle I \rangle^2$
for 4 high-energy bins  and for different flux thresholds in units of the Crab flux, and  for
 12 additional energy bins in the range 500~MeV -- 200~GeV as in Fig.~\ref{fig:Cptwo}.}
\label{fig:anisadcta}
\medskip
\end{figure}

\begin{figure}
\centering
\includegraphics[width=0.48\textwidth]{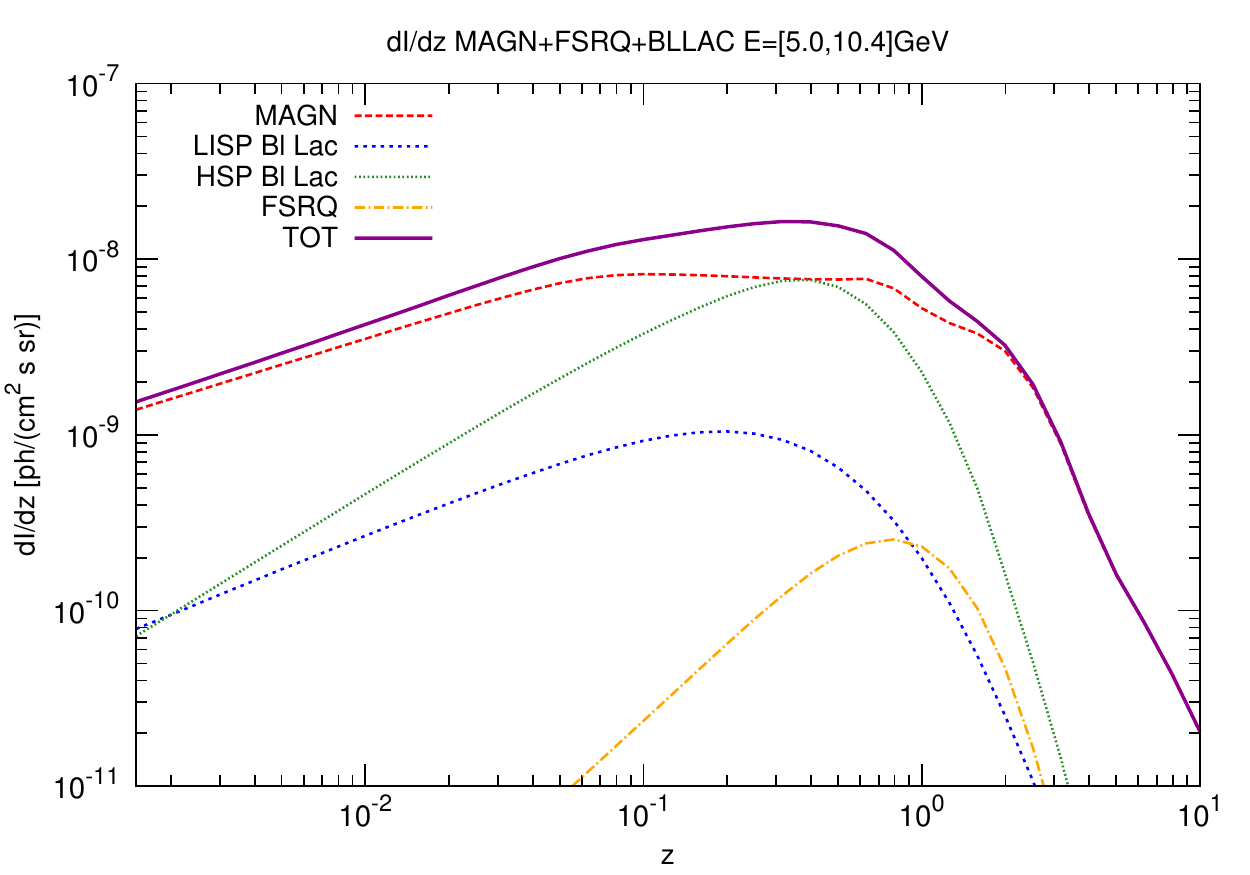}
\includegraphics[width=0.48\textwidth]{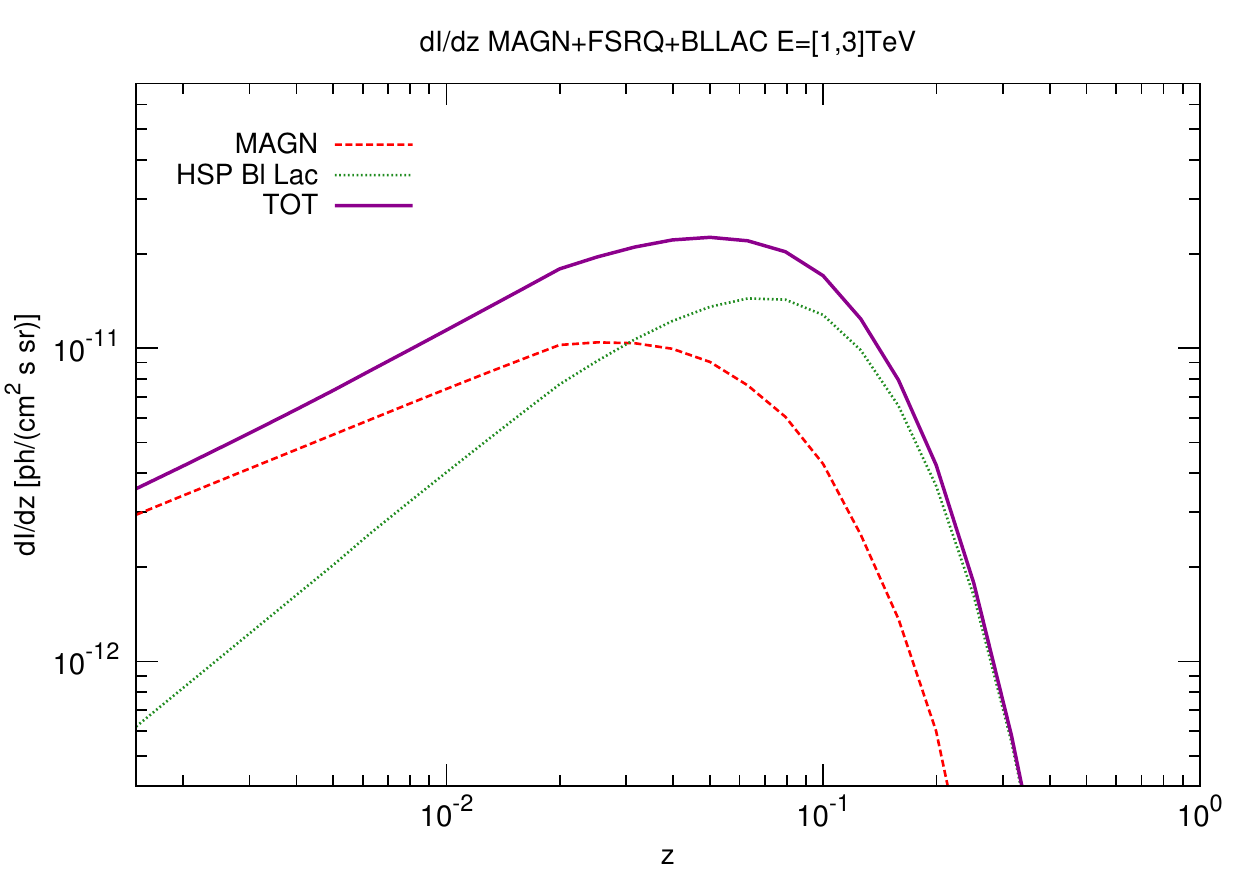}
\includegraphics[width=0.48\textwidth]{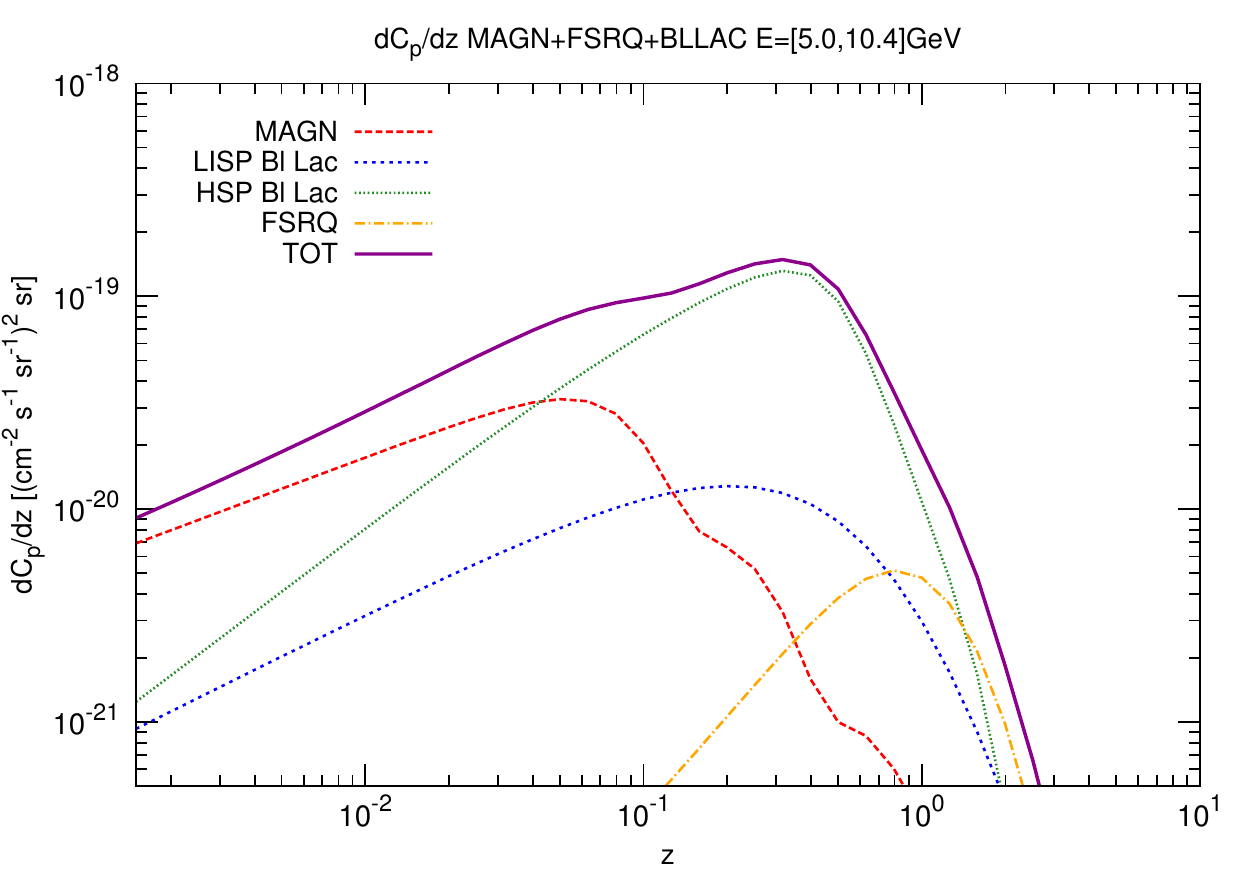}
\includegraphics[width=0.48\textwidth]{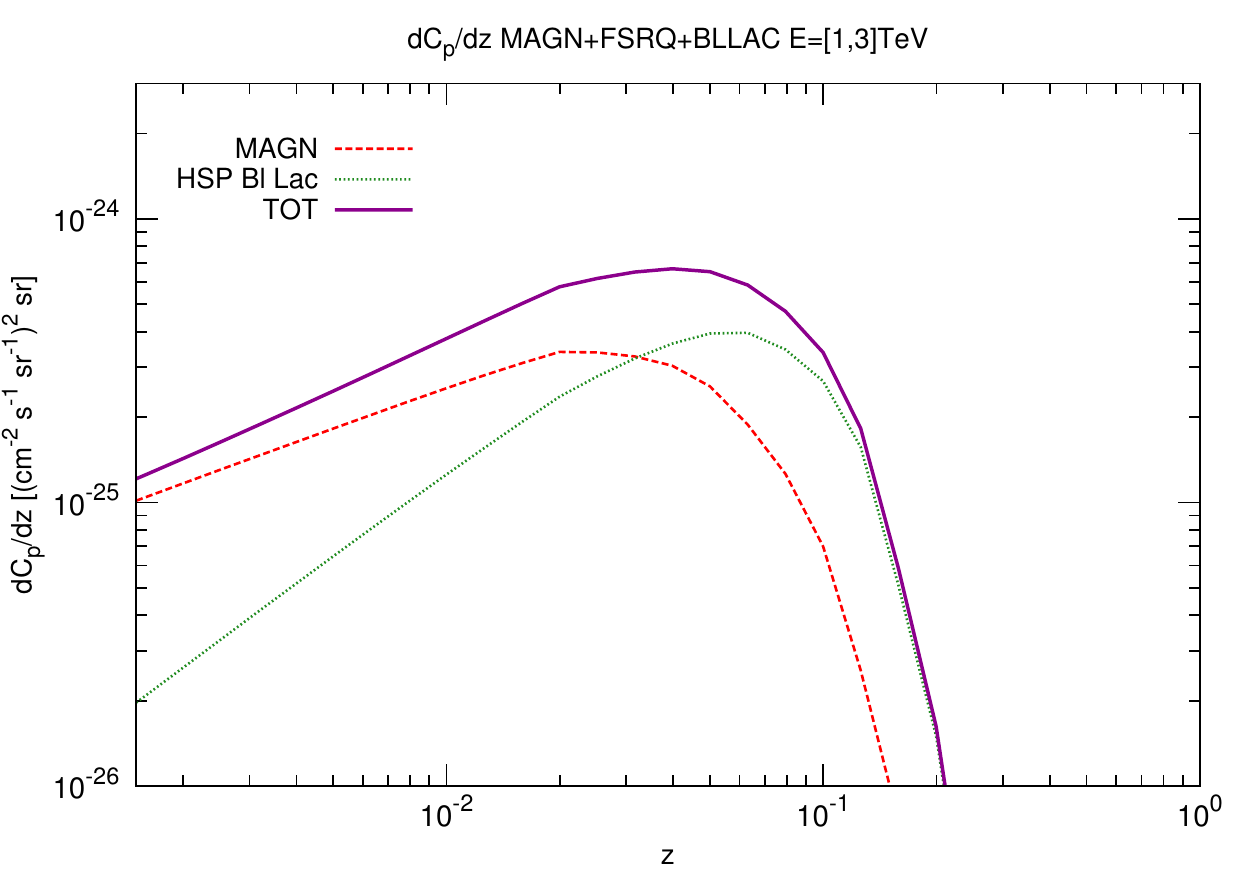}
\caption{IGRB intensity distribution $dI/dz$ and anisotropy distribution $dC_P/dz$
as function of redshift $z$. The curves show the separate
contributions from each component and the total.
The left column shows the intensity and anisotropy in the 5-10.4 GeV energy bin,
while the right column refers to the 1-3 TeV energy bin and for a 
detection threshold of 10 mCrab.}
\label{fig:dIdzdCpdz}
\medskip
\end{figure}

Finally,  we show in Fig.~\ref{fig:anisadcta} also the \emph{fluctuation angular power}
as function of energy, i.e.  the quantity $C_{\rm P}/\langle I \rangle^2$, where $\langle I \rangle$ is the mean intensity of the emission, which is
 another useful measure of the anisotropy properties of the unresolved emission~\cite{2012PhRvD..85h3007A,2009PhRvL.102x1301S}.
In particular, for a given threshold  flux and  a given energy bin, $C_{\rm P}$ and $I$
are given by Eqs.~\ref{Cpdef} and \ref{Idef}, which means that
$C_{\rm P}/\langle I \rangle^2$ could have a non-trivial behavior as function of the threshold  flux
since both $C_{\rm P}$ and $I$ vary with it.
It can be seen that below 100 GeV the $C_{\rm P}/\langle I \rangle^2$ curve is approximately
flat around $\sim 10^{-4}$.
This value is in agreement with the value of $\sim 10^{-5}$ measured
with the {\it Fermi}-LAT in~\cite{2012PhRvD..85h3007A}, after accounting for the fact 
that   the  $\langle I \rangle$  used in~\cite{2012PhRvD..85h3007A}  to calculate  $C_{\rm P}/\langle I \rangle^2$ is the 
average \emph{high-latitude} diffuse emission rather than the true
IGRB, and the former is roughly a factor of 3 larger than the true
IGRB due to the Galactic diffuse emission contribution.
Above 100 GeV $C_{\rm P}/\langle I \rangle^2$ has a clear upturn.
This is due to the partial attenuation of the extragalactic $\gamma$-ray photons
by the extra-galactic background light (EBL) due to $e^+e^-$pair production.
This effect is taken into account in our calculations
using the EBL model of~\cite{2010ApJ...712..238F}.
As cross-check we also verified that our results do not change significantly when using the 
alternative EBL attenuation models of \cite{Franceschini:2008tp,Gilmore:2011ks,Dominguez:2010bv}.
All the above models are, indeed, in rough agreement
among each other  and  with the observed  high-energy spectra
of blazars as seen by the {\it Fermi}-LAT and HESS~\cite{Ackermann:2012sza,Abramowski:2012ry}.
In practice, the EBL attenuation causes a steepening of the IGRB spectrum by
absorbing the high-redshift, high-energy contribution to the IGRB\@.
The high-redshift part of the IGRB is also typically the most
isotropic since, due to volume effects,  is made by many distant (and thus faint)
sources, so that it gives only a minor contribution to $C_{\rm P}$.
For this reason,  the anisotropy $C_{\rm P}$
is not significantly affected by the EBL, in contrast to the intensity $\langle I \rangle$ which is more strongly affected,
so that the ratio $C_{\rm P}/\langle I \rangle^2$ increases with energy.
This effect is illustrated in Fig.~\ref{fig:dIdzdCpdz}, where the
IGRB intensity distribution $dI/dz$ and anisotropy distribution $dC_P/dz$ are shown
as function of redshift $z$ for the two energy bins 5-10.4 GeV and 1-3 TeV.
At energies of few GeVs, it can be seen indeed that while the IGRB intensity is produced 
at redshifts up to 3-4, the anisotropy originates basically from more nearby redshifts of $z\lesssim$1.
At TeV energies, instead, where the effect  of the EBL absorption is significant,
 both the intensity and anisotropy originates from nearby redshifts $z\lesssim$ 0.1.

Using $C_{\rm P}/\langle I \rangle^2$ we can also compare our results to the sensitivity study of IACTs to anisotropies.
In~\cite{Ripken:2012db} anisotropies from both dark matter and astrophysical sources
have been studied in the two energy ranges $E>$ 100 GeV and $E>$ 300 GeV.  
It is shown that CTA should be sensitive to 
 anisotropies in the astrophysical component down to a value as low as $10^{-5}$. 
A value of $\sim 10^{-3}$ as predicted by our analysis above 100 GeV should
thus be easily detectable with CTA\@. Indeed, a value as high as $\sim 10^{-3}$
could already be within the reach of present day IACTs.
On the other hand, a value of $\sim 10^{-3}$ could present a challenge
for dark matter searches in the TeV range since in~\cite{Ripken:2012db} an astrophysical background of $\sim 10^{-5}$
was assumed. 
A re-analysis of the sensitivity of IACTs to dark-matter--induced anisotropy considering 
the present up-to-date models of the radio-loud AGN population  will be
required to accurately understand the potential of these instruments to use anisotropies to search for dark matter.

\subsection{Discussion}

As already noted above, the anisotropy at CTA energies, above 100 GeV, is dominated
by HSP BL Lacs and  MAGNs.  
The predicted intensity and anisotropy depends
on the SED of these populations which, at these high energies, is
still affected  by various uncertainties due to the scarcity of 
sources observed.
For example, the effect of the assumed energy cutoff of HSPs of 910 GeV
is clearly visible in Fig.~\ref{fig:aniscta}, with the anisotropy in the
last bin being considerably suppressed.
On the other hand, considering the 1-$\sigma$ uncertainties in \cite{DiMauro:2013zfa},
the energy cutoff can be in principle  in the range 460-2010 GeV.
Variation of the cutoff within this range will
correspondingly affect the level of anisotropy, especially in the 2 highest-energy bins.
For MAGNs a cutoff is not observed, while,
instead, at least in the case of Cen A, a hardening above 
$\sim$10 GeV is seen \cite{Sahakyan:2013eha}.
If this is a general feature of MAGNs, a corresponding
enhancement in the anisotropy from MAGNs at high energies  should be expected.
Again, due to very limited number of sources
detected at TeV energies, a robust conclusion is not possible.
Finally, some further classes of TeV  sources might be present.
For example,
extreme HBLs, sometimes dubbed ultra-high-frequency-peaked BL Lac \cite{Senturk:2013pa,Lefa:2011xh}, 
are BL Lacs with the synchrotron peak at 
MeV energies and they show at IACTs energies a hardening of the spectra.
As discussed in  \cite{DiMauro:2013zfa}, 20\% of the HSP sample does not show a cut off in the spectra and this sample could be
part of this extreme class of BL Lacs. Although the sample is not homogenous,
it can be reasonably assumed that ultra-high-frequency-peaked BL Lacs represent around 20\% of HSP population. 
This class of BL Lacs could thus contribute to enhance the anisotropy at TeV energies.  
Clearly, to fully address this issue, larger statistics of BL Lacs sources at TeV energies are required.
The situation should become more clear in the following years, as more sources will be hopefully detected.

Finally, we add some consideration on the effective detectability of
the IGRB intensity and anisotropy with CTA.
This issue is discussed, to some extent, in \cite{Ripken:2012db}.
Due to the large amount of CR background, especially electrons, in IACT 
observations,  the signal-to-noise ratio for the IGRB
intensity measurement is quite low
and the prospects for a measurement are limited. 
On the contrary,
somewhat counter-intuitively, detection of IGRB anisotropies might
be feasible  since the CR background
is not expected to exhibit small scale anisotropies
giving a larger signal-to-noise.
In \cite{Ripken:2012db}, the prospects for IGRB anisotropy
detection with CTA have been investigated in a simplified
setup and found to be promising (see also discussion in the previous section).
A deeper study using detailed simulations of  CTA 
observations will help to spot potential unseen
problematics  and eventually to confirm the possibility of anisotropy observations with CTA.

\section{Conclusions}
\label{sec:conc}

In this work we have studied the contribution of various classes of radio-loud AGN to the intensity and anisotropy of the IGRB\@.
We have used up-to-date phenomenological models for the $\gamma$-ray luminosity functions of these populations
to predict the $\gamma$-ray flux and anisotropy as functions of the energy.
We have found that the entirety of the IGRB intensity and anisotropy as measured by the {\it Fermi}-LAT
can be explained by radio-loud AGN, with  MAGN providing the 
bulk of the measured intensity but a low anisotropy, while high-synchrotron-peaked BL Lac objects 
give the main contribution to the anisotropy. 
Nonetheless,  the predicted intensity from MAGN  still suffers from large uncertainties,
which are difficult to reduce even with the use of  anisotropy information,
given the low level of anisotropy predicted from this population.

Within these uncertainties, there may be still some room for 
further $\gamma$-ray populations, provided that their anisotropy remains low in order not to exceed 
the measurements, which are already fairly saturated by the HSP BL Lac contribution.
In this respect, for example, star-forming galaxies  could still provide a substantial contribution to the
low-energy side of the IGRB, while exhibiting a low level of anisotropy and 
 keeping a self-consistent picture in agreement with the observed IGRB intensity  and anisotropy.

Upcoming analyses of 5 years of {\it Fermi}-LAT data below about 100 GeV and forthcoming observations with
the Cherenkov Telescope Array above 100 GeV will provide valuable tests of 
the AGN scenario depicted in the present analysis, and will contribute significantly to improving our knowledge of  high-energy $\gamma$-ray emitters.

\acknowledgments

We thank  Pasquale D. Serpico and Hannes Zechlin for useful discussions and a careful reading of the manuscript. 
This work is supported by the research grant {\sl TAsP (Theoretical Astroparticle Physics)}
funded by the Istituto Nazionale di Fisica Nucleare (INFN), by the  {\sl Strategic Research Grant: Origin and Detection of Galactic and Extragalactic Cosmic Rays} funded by Torino University and Compagnia di San Paolo, by the Spanish MINECO under grants FPA2011-22975 and MULTIDARK CSD2009-00064 (Consolider-Ingenio 2010 Programme), by Prometeo/2009/091 (Generalitat Valenciana), and by the EU ITN UNILHC PITN-GA-2009-237920. 
At LAPTh this activity was supported by the Labex grant ENIGMASS.  JSG acknowledges support from NASA through Einstein Postdoctoral Fellowship grant PF1-120089 awarded by the Chandra X-ray Center, which is operated by the Smithsonian Astrophysical Observatory for NASA under contract NAS8-03060.

\bibliographystyle{JHEP}
\bibliography{anisotropy_v5}

\end{document}